\documentclass[sigchi, authorversion]{acmart}
\AtBeginDocument{%
  \providecommand\BibTeX{{%
    \normalfont B\kern-0.5em{\scshape i\kern-0.25em b}\kern-0.8em\TeX}}}

\copyrightyear{2023}
\acmYear{2023}
\setcopyright{rightsretained}
\acmConference[DIS '23]{Designing Interactive Systems Conference}{July 10--14, 2023}{Pittsburgh, PA, USA}
\acmBooktitle{Designing Interactive Systems Conference (DIS '23), July 10--14, 2023, Pittsburgh, PA, USA}
\acmDOI{10.1145/3563657.3595964}
\acmISBN{978-1-4503-9893-0/23/07}

\usepackage{color, colortbl}
\usepackage{multirow}
\usepackage{subfig}
\usepackage{hyperref}
\usepackage[font=small,labelfont=bf]{caption}
\begin{document}

\title{Defending Against the Dark Arts: Recognising Dark Patterns in Social Media}

\author{Thomas Mildner}
\affiliation{%
  \institution{University of Bremen}
  \city{Bremen}
  \country{Germany}
}
\email{mildner@uni-bremen.de}
\orcid{0000-0002-1712-0741}

\author{Merle Freye}
\affiliation{%
  \institution{University of Bremen}
  \city{Bremen}
  \country{Germany}
}
\email{mfreye@uni-bremen.de}
\orcid{0000-0001-9895-9997}

\author{Gian-Luca Savino}
\affiliation{%
  \institution{University of St.Gallen}
  \city{St.Gallen}
  \country{Switzerland}
}
\email{gian-luca.savino@unisg.ch}
\orcid{0000-0002-1233-234X}

\author{Philip R. Doyle}
\affiliation{%
  \institution{University College Dublin}
  \city{Dublin}
  \country{Ireland}
}
 \email{philip.doyle1@ucdconnect.ie}
\orcid{0000-0002-2686-8962}

\author{Benjamin R. Cowan}
\affiliation{%
  \institution{University College Dublin}
  \city{Dublin}
  \country{Ireland}
}
\email{benjamin.cowan@ucd.ie}
\orcid{0000-0002-8595-8132}

\author{Rainer Malaka}
\affiliation{%
  \institution{University of Bremen}
  \city{Bremen}
  \country{Germany}
}
\email{malaka@tzi.de}
\orcid{0000-0001-6463-4828}

\renewcommand{\shortauthors}{Mildner et al.}

\begin{abstract}
Interest in unethical user interfaces has grown in HCI over recent years, with researchers identifying malicious design strategies referred to as ``dark patterns''. While such strategies have been described in numerous domains, we lack a thorough understanding of how they operate in social networking services (SNSs). Pivoting towards regulations against such practices, we address this gap by offering novel insights into the types of dark patterns deployed in SNSs and people's ability to recognise them across four widely used mobile SNS applications. Following a cognitive walkthrough, experts ($N=6$) could identify instances of dark patterns in all four SNSs, including co-occurrences. Based on the results, we designed a novel rating procedure for evaluating the malice of interfaces. Our evaluation shows that regular users ($N=193$) could differentiate between interfaces featuring dark patterns and those without. Such rating procedures could support policymakers' current moves to regulate deceptive and manipulative designs in online interfaces.

\end{abstract}



\maketitle


\section{Introduction}
Among HCI researchers, interest in the ethical implications of how technology is designed has seen a noticeable increase over recent years. One of the more widely known topics within this work is research that focuses on unethical design strategies, referred to as ``dark patterns''. Cataloguing instances of dark patterns has led to a growing collection of interface artefacts that negatively affect users' ability to make informed decisions. A common example can be seen in cookie-consent banners that often visually elevate options allowing the tracking and storing of users' data over alternatives to denying such functionalities. Originating in e-commerce \cite{brignull_deceptive_nodate, mathur2019}, and other online websites~\cite{brignull_deceptive_nodate, gray2018}, dark patterns describe design strategies that coerce, steer, or obfuscate users into unfavourable actions that they may not have taken if they were fully informed~\cite{mathur2021}. Today, related work has identified a multitude of designs that fit this definition, including digital games~\cite{zagal_dark_2013}, social networking sites (SNS)~\cite{mildner_ethical_2021, gunawan_comparative_2021, habib_identifying_2022, schaffner_understanding_2022}, and mobile applications~\cite{digeronimo2020, BongardBlanchy2021, gray2018}.

The adverse effects of dark patterns have drawn the attention of regulators worldwide. Examples aimed at better protecting users’ privacy and autonomy can be seen in the California Consumer Privacy Act CCPA~\cite{ccpa_2018} or the Digital Service Act (DSA) of the European Union~\cite{eu_commission_2022_dsa}. Regardless of the national background, regulating dark patterns faces common challenges, such as a missing taxonomy, the rapid development of new dark patterns, and difficulty identifying dark patterns that require legal interventions. We see that findings from human-computer interaction (HCI) can support the legal discussion and legislative efforts~\cite{Gray2021} in developing a taxonomy and providing the right tools to assess and regulate dark patterns. Therefore, it is crucial that research advances our understanding of the implications of dark patterns in as many domains as possible to enable regulators and legislators to create effective measures to protect users. 

In this work, we take steps towards achieving this goal by (1) analysing the ability of experts and regular users of social media to identify dark patterns based on established definitions thereof and by (2) studying an alternative approach to classify interfaces based on high-level characteristics proposed by Mathur et al.~\cite{mathur2019, mathur2021} to approach an easier evaluation. As this is a relatively new research area, knowledge about how people perceive dark patterns is still limited, with a handful of studies exploring this particular aspect of the topic~\cite{digeronimo2020, BongardBlanchy2021, maier_dark_2020}. In light of initial moves towards regulation and increased attention in the scientific literature, this work reflects on the current state of the dark pattern research, investigates how applicable current taxonomies are in domains in which they were not first established, and whether current definitions can be utilised as evaluation tools. Before conducting this research, we collected 69 types of dark patterns from eight papers~\cite{zagal_dark_2013, brignull_deceptive_nodate, conti_malicious_2010, greenberg_dark_nodate, bosch2016, gray_ethical_2019, Gray2020a, mathur2019}, further included in Mathur et al.'s~\cite{mathur2021} literature review. While we are aware that recent work have updated the overall corpus of dark patterns~\cite{mildner2022, gunawan_comparative_2021}, which we could not include in our studies, the focus of this research is to aim for a simplified recognition tool to aid policy-makers' and regulators' efforts. For this endeavor, we turn towards SNSs as we still lack certain insights about how malicious interfaces manifest in this context. Additionally, the omnipresent nature of SNSs affords constant investigation as research repeatedly highlights negative effects posed on their users' well-being~\cite{beyens2020effect, ahn2013social}. Aiming to aid regulatory efforts, we address these research gaps based on two research questions:

\begin{itemize}
    \item[\textbf{RQ1}] Can dark patterns taxonomies be used by experts to identify and recognise instances in SNSs?
    \item[\textbf{RQ2}] Are regular SNS users able to differentiate between interfaces with and without dark patterns?
\end{itemize}

We answer these questions through two studies. In the first, we conducted cognitive walkthroughs with six HCI researchers aimed at investigating whether current dark pattern taxonomies can be used to assess and identify dark patterns in novel interfaces. The four SNSs included in the study were Facebook, Instagram, TikTok, and Twitter. In a second study, we conducted an online survey to learn about the recognisability of dark patterns by regular SNS users. In contrast to the first study, we did not provide participants of the second study with the complete corpus of dark pattern research but instead relied on five questions adopting Mathur et al.'s~\cite{mathur2019,mathur2021} high-level dark pattern characteristics with the aim of assessing the malice of a particular interface design. While this hinders an immediate comparison between both studies, our evaluation of this alternative process shows that regular users are able to generally recognise dark patterns. Conclusively, dark patterns were not rated to be very malicious (using Mathur et al.’s~\cite{mathur2019,mathur2021} five high-level characteristics) but participants were able to successfully discern dark patterns from a selection of interface screenshots collected from Study 1, that either did or did not contain them. We also propose that a similar approach, one that is not fundamentally linked to specific examples of dark pattern design, could introduce more flexibility and practicality into current legislation processes and would better future-proof legislative efforts aiding the protection of users.


\section{Related Work}
In this section, we will approach relevant research to identify, recognise, and regulate dark patterns from two directions. We will begin by establishing a taxonomy of dark pattern types resulting from the collaborative effort of prior research. This taxonomy is later used in our first study. Afterwards, we highlight work studying the perception and recognition of dark patterns, a necessary step towards successful regulation. We then outline the form of current approaches and strategies in the final paragraphs of this section.

\subsection{Dark Pattern Taxonomy}
\begin{table*}[t!]
\resizebox{1\textwidth}{!}{%
\begin{tabular}{p{0.125\linewidth}p{0.125\linewidth}p{0.125\linewidth}p{0.125\linewidth}p{0.125\linewidth}p{0.125\linewidth}p{0.125\linewidth}p{0.125\linewidth}}
\toprule
\multicolumn{1}{c}{\begin{tabular}[c]{@{}c@{}}\LARGE\textbf{Brignull}\\ \small{2010 ~\cite{brignull_deceptive_nodate}}\end{tabular}} & 
\multicolumn{1}{c}{\begin{tabular}[c]{@{}c@{}}\LARGE\textbf{Conti \& Sobiesk}\\ \small{2010~\cite{conti_malicious_2010}}\end{tabular}} &
\multicolumn{1}{c}{\begin{tabular}[c]{@{}c@{}}\LARGE\textbf{Zagal et al.}\\ \small{2013~\cite{zagal_dark_2013}}\end{tabular}} & 
\multicolumn{1}{c}{\begin{tabular}[c]{@{}c@{}}\LARGE\textbf{Greenberg et al.}\\ \small{2014~\cite{greenberg_dark_nodate}}\end{tabular}} & 
\multicolumn{1}{c}{\begin{tabular}[c]{@{}c@{}}\LARGE\textbf{Bösch et al.}\\ \small{2016~\cite{bosch2016}}\end{tabular}} & 
\multicolumn{1}{c}{\begin{tabular}[c]{@{}c@{}}\LARGE\textbf{Gray et al.}\\ \small{2018~\cite{gray2018}}\end{tabular}} & 
\multicolumn{1}{c}{\begin{tabular}[c]{@{}c@{}}\LARGE\textbf{Gray et al.}\\ \small{2020~\cite{Gray2020a}}\end{tabular}} & 
\multicolumn{1}{c}{\begin{tabular}[c]{@{}c@{}}\LARGE\textbf{Mathur et al.}\\ \small{2019~\cite{mathur2019}}\end{tabular}} 
\\ \midrule

\multicolumn{1}{c|}{\begin{tabular}[t]{l}   \medskip\large\textit{· Trick Questions} \\
                                            \medskip\large\textit{· Sneak Into Basket} \\
                                            \medskip\large\textit{· Roach Motel} \\
                                            \medskip\large\textit{· Privacy Zuckering} \\
                                            \medskip\large\textit{· Confirmshaming} \\
                                            \medskip\large\textit{· Disguised Ads} \\
                                            \medskip\large\textit{\begin{tabular}[l]{@{}l@{}}· Price Comparison\\\hspace{5pt}Prevention\end{tabular}} \\
                                            \medskip\large\textit{· Misdirection} \\
                                            \medskip\large\textit{· Hidden Costs} \\
                                            \medskip\large\textit{· Bait and Switch} \\
                                            \medskip\large\textit{· Forced Continuity} \\
                                            \medskip\large\textit{· Friend Spam} \\
\end{tabular}} &                                       
\multicolumn{1}{c|}{\begin{tabular}[t]{l}   \medskip\large\textit{· Coercion} \\ 
                                            \medskip\large\textit{· Distraction} \\
                                            \medskip\large\textit{· Forced Work} \\
                                            \medskip\large\textit{\begin{tabular}[l]{@{}l@{}}· Manipulating\\\hspace{5pt}Navigation\end{tabular}} \\
                                            \medskip\large\textit{\begin{tabular}[l]{@{}l@{}}· Restricting\\\hspace{5pt}Functionality\end{tabular}} \\
                                            \medskip\large\textit{· Trick} \\
                                            \medskip\large\textit{· Confusion} \\
                                            \medskip\large\textit{· Exploiting Errors} \\
                                            \medskip\large\textit{· Interruption} \\
                                            \medskip\large\textit{· Obfuscation} \\
                                            \medskip\large\textit{· Shock} \\
\end{tabular}} &                                                           
\multicolumn{1}{c|}{\begin{tabular}[t]{l}   \medskip\large\textit{· Grinding} \\ 
                                            \medskip\large\textit{· Impersonation} \\
                                            \medskip\large\textit{· Monetized Rivalries} \\
                                            \medskip\large\textit{· Pay to Skip} \\
                                            \medskip\large\textit{\begin{tabular}[l]{@{}l@{}}· Playing by\\\hspace{5pt}Appointment\end{tabular}}\\
                                            \medskip\large\textit{\begin{tabular}[l]{@{}l@{}}· Pre-Delivered \\\hspace{5pt}Content\end{tabular}}\\
                                            \medskip\large\textit{\begin{tabular}[l]{@{}l@{}}· Social Pyramid\\\hspace{5pt}Schemes\end{tabular}}\\
\end{tabular}} &                                                                
\multicolumn{1}{c|}{\begin{tabular}[t]{l}   \medskip\large\textit{· Attention Grabber} \\ 
                                            \medskip\large\textit{· Bait and Switch} \\ 
                                            \medskip\large\textit{\begin{tabular}[l]{@{}l@{}}· The Social Network\\\hspace{5pt}Of Proxemic Contracts\\\hspace{5pt}Or Unintended\\\hspace{5pt}Relationships\end{tabular}}\\ 
                                            \medskip\large\textit{· Captive Audience} \\ 
                                            \medskip\large\textit{· We Never Forget} \\ 
                                            \medskip\large\textit{\begin{tabular}[l]{@{}l@{}}· Disguised Data\\\hspace{5pt}Collection\end{tabular}}\\ 
                                            \medskip\large\textit{\begin{tabular}[l]{@{}l@{}}· Making Personal\\\hspace{5pt}Information Public\end{tabular}}\\ 
                                            \medskip\large\textit{· The Milk Factor} \\ 

\end{tabular}} &                                                              
\multicolumn{1}{c|}{\begin{tabular}[t]{l}   \medskip\large\textit{· Privacy Zuckering} \\ 
                                            \medskip\large\textit{\begin{tabular}[l]{@{}l@{}}· Hidden Legalese\\\hspace{5pt}Stipulations\end{tabular}} \\
                                            \medskip\large\textit{· Shadow User Profiles} \\
                                            \medskip\large\textit{· Bad Defaults} \\
                                            \medskip\large\textit{· Immortal Accounts} \\
                                            \medskip\large\textit{· Information Milking} \\
                                            \medskip\large\textit{· Forced Registration} \\
                                            \medskip\large\textit{\begin{tabular}[l]{@{}l@{}}· Address Book\\\hspace{5pt}Leeching\end{tabular}} \\

\end{tabular}} &                                                           
\multicolumn{1}{c|}{\begin{tabular}[t]{l}   \medskip\large\textit{· Nagging} \\ 
                                            \medskip\large\textit{· Obstruction} \\
                                            \medskip\large\textit{· Sneaking} \\
                                            \medskip\large\textit{· Interface Interference} \\
                                            \medskip\large\textit{· Forced Action} \\

\end{tabular}} &                                                        
\multicolumn{1}{c|}{\begin{tabular}[t]{l}   \medskip\large\textit{· Automating the User} \\ 
                                            \medskip\large\textit{· Two-Faced} \\
                                            \medskip\large\textit{· Controlling} \\
                                            \medskip\large\textit{· Entrapping} \\
                                            \medskip\large\textit{· Nickling-And-Diming} \\
                                            \medskip\large\textit{· Misrepresenting} \\

\end{tabular}} &  
\multicolumn{1}{l}{\begin{tabular}[t]{l}    \medskip\large\textit{· Countdown Timers} \\ 
                                            \medskip\large\textit{\begin{tabular}[l]{@{}c@{}}· Limited-time\\\hspace{5pt}Messages\end{tabular}} \\
                                            \medskip\large\textit{\begin{tabular}[l]{@{}c@{}}· High-demand\\\hspace{5pt}Messages\end{tabular}} \\
                                            \medskip\large\textit{· Activity Notifications} \\
                                            \medskip\large\textit{· Confirmshaming} \\
                                             \medskip\large\textit{\begin{tabular}[l]{@{}c@{}}· Testimonials\\\hspace{5pt}of Uncertain\\\hspace{-18pt}Origins\end{tabular}} \\
                                            \medskip\large\textit{· Hard to Cancel} \\
                                            \medskip\large\textit{· Visual Interference} \\
                                            \medskip\large\textit{· Low-stock Messages} \\
                                            \medskip\large\textit{· Hidden Subscriptions} \\
                                            \medskip\large\textit{· Pressured Selling} \\
                                            \medskip\large\textit{· Forced Enrollment} \\
\end{tabular}} 

\\ \bottomrule
\end{tabular}
}
\caption{This table shows 69 types of dark patterns described in eight related works. Columns are in chronological order in which these works were published.}
\Description[This table shows types of dark pattern described in prior research.]{This table presents a total of 69 types of dark pattern divided by their original publication over 8 columns. These columns are as follows: 
1. Brignull et al. 
· Trick Questions
· Sneak Into Basket
· Roach Motel
· Privacy Zuckering
· Confirmshaming
· Disguised Ads
· Price Comparison
Prevention
· Misdirection
· Hidden Costs
· Bait and Switch
· Forced Continuity
· Friend Spam 

2. Conti & Sobiesk
· Coercion
· Distraction
· Forced Work
· Manipulating
Navigation
· Restricting
Functionality
· Trick
· Confusion
· Exploiting Errors
· Interruption
· Obfuscation
· Shock

3. Zagal et al.
· Grinding
· Impersonation
· Monetized Rivalries
· Pay to Skip
· Playing by
Appointment
· Pre-Delivered
Content
· Social Pyramid
Schemes

4. Greenberg et al.
· Attention Grabber
· Bait and Switch
· The Social Network
Of Proxemic Contracts
Or Unintended
Relationships
· Captive Audience
· We Never Forget
· Disguised Data
Collection
· Making Personal
Information Public
· The Milk Factor

5. Bösch et al.
· Privacy Zuckering
· Hidden Legalese
Stipulations
· Shadow User Profiles
· Bad Defaults
· Immortal Accounts
· Information Milking
· Forced Registration
· Address Book
Leeching

6. Gray et al. 
· Nagging
· Obstruction
· Sneaking
· Interface Interference
· Forced Action

7. Gray et al.
· Automating the User
· Two-Faced
· Controlling
· Entrapping
· Nickling-And-Diming
· Misrepresenting

8. Mathur et al.
· Countdown Timers
· Limited-time
Messages
· High-demand
Messages
· Activity Notifications
· Confirmshaming
· Testimonials
of Uncertain
Origins
· Hard to Cancel
· Visual Interference
· Low-stock Messages
· Hidden Subscriptions
· Pressured Selling
· Forced Enrollment
}
\label{tab:dp_taxonomy}
\end{table*}

Here, we attempt to provide a relatively comprehensive overview of the current dark patterns landscape. 
To provide a summary of the taxonomy used in our studies, Table~\ref{tab:dp_taxonomy} presents key contributions taken from Mathur et al.'s~\cite{mathur2021} earlier review on dark pattern literature. As we deem it important for our studies that the definitions for dark patterns should be the result of empirical research, we decided to limit the scope for the eight academic contributions part of Mathur et al.'s literature review~\cite{mathur2021}. Although more holistic guidelines exist, these are not included as they tend not to provide enough empirical evidence in their definitions. This left eight academic works that met our criteria, which collectively presented 69 different types of dark patterns that are outlined below in chronological order. Brignull~\cite{brignull_deceptive_nodate}, who first coined the term dark pattern, initialised the current body of work with twelve types that concern online design strategies. In a similar effort, Conti and Sobiesk~\cite{conti_malicious_2010} defined eleven types of malicious strategies based on a one-year data collection. Although their work was published before the term dark pattern gained the recognition it sees today, we refer to their results as dark patterns for the sake of conciseness. Offering seven game-specific dark patterns, Zagal et al.~\cite{zagal_dark_2013} studied tricks used in that industry to create, for example, competition or disparate treatment through unethical practices. In another work, Greenberg et al.~\cite{greenberg_dark_nodate} were interested in the possible exploitation of spatial factors when discussing dark patterns through the lens of proxemic theory. The result introduces eight types of proxemic dark patterns like speculative technologies targeting users with specific advertisements using public displays. Closely related to the Privacy by Design concept~\cite{hustinx_privacy_2010}, and thus particularly interesting for our research, Bösch et al.~\cite{bosch2016} collected eight types of dark patterns enveloping schemes that target data collection and limitations of users' agency to customise their personal preferences.

Taking a different approach, Gray et al.~\cite{gray2018} looked to investigate how dark patterns are created in the first place. Here, researchers analysed an image-based corpus of potential types of dark patterns using a qualitative approach while relying on Brignull's original taxonomy. They define five types of dark patterns that practitioners engage in when developing manipulative designs. Following this research, Gray et al.~\cite{gray_ethical_2019} applied content analysis on 4775 user-generated posts collected from the Reddit sub-forum \textit{r/assholedesign}. Their result provides six properties ``asshole designers'' subscribe to. Interested in the number of web services embedding dark patterns, Mathur et al.~\cite{mathur2019} applied hierarchical clustering to identify that 11\% of shopping websites employ text-based dark patterns based on a collection of more than 11k samples. Evaluation of their data generated twelve dark patterns embedded in shopping websites.

These works bring together 69 types of dark patterns. Noticeably, various domains have been investigated, widening our understanding of these strategies' origins. However, there is currently a potentially important gap regarding SNS-related platforms like Facebook, Instagram, TikTok, and Twitter -- platforms that many people interact with frequently in their day-to-day lives. A growing body of research already illustrates problems with users accurately recollecting the amount of time they spend on SNSs and the frequency in which they use these services~\cite{junco_comparing_2013, schoenebeck_giving_2014, ernala_how_2020}. Concerns are also growing regarding alarming implications SNSs have on their users' well-being~\cite{wang2011, wang_effects_2014, beyens2020effect, shakya_association_2017}. Filling this gap, the research presented here considers the current discourse to review the presence of these described dark patterns in four major SNS platforms.

\subsection{Perceiving Dark Patterns}
Interested in the cognitive biases dark patterns exploit, Mathur et al.~\cite{mathur2019} analysed their dark patterns further and recognised five common characteristics in which these dark patterns operate: \textit{asymmetric}; \textit{restrictive}; \textit{covert}; \textit{deceptive}; and \textit{information hiding}. In a follow-up effort, Mathur et al~\cite{mathur2021} applied these characteristics to prior dark pattern taxonomies while extending the framework to include a sixth characteristic named \textit{disparate treatment}. Collectively, this framework promises an alternative and interesting tool to study dark patterns. To test its utility outside its original scope, our research applies this framework to recognise dark patterns in SNSs.
Instead of focusing entirely on the identification of dark patterns, a multitude of works considers end-users' perspectives of dark patterns. In this sense, Di Geronimo et al.~\cite{digeronimo2020} sampled 240 popular applications from the Google Playstore and analysed each for contained dark patterns based on Gray et al.’s~\cite{gray2018} taxonomy. Based on 10-minute cognitive walkthroughs, their results indicate that 95\% of tested applications yield dark patterns. An ensuing online survey revealed that the majority of users fail to discern Dark Patterns in 30-second video recordings of mobile applications. However, their ability to identify harmful designs improves when educated on the subject. In line with prior research, including Maier and Harr's~\cite{maier_dark_2020} confirmation of users' difficulty to recognise dark patterns~\cite{maier_dark_2020}, Bongard-Blanchy et al.~\cite{BongardBlanchy2021} reinforce these implications through their online survey studying participants' ability to recognise dark patterns. Studying the effects browser modalities have on the number of dark patterns users are faced with, Gunawan et al~\cite{gunawan_comparative_2021} conducted a thematic analysis on recordings of various online services. Their work describes twelve previously not described dark patterns, including \textit{extraneous badges} that describe nudging interface elements, like coloured circles, which provoke immediate interaction. Trying to understand Facebook users' control over ad-related settings, Habib et al.~\cite{habib_identifying_2022} demonstrate that the SNS does not meet users' preferred requirements. Considering dark patterns in their work, the authors discuss problematic interface structures limiting users' agency to choose settings efficiently and to their liking. This limitation is further discussed by Schaffner et al.~\cite{schaffner_understanding_2022}, who demonstrate difficulties for users to successfully delete their accounts across 20 SNSs. Their success rate was additionally impacted by the modality in which a particular SNS is accessed.

Investigating persuasive designs, Utz et al.~\cite{utz_informed_2019} demonstrate how nudging interfaces can shift users' decisions towards a preset goal. In a similar vein, Graßl et al.~\cite{grasl_dark_2021} showed evidence that nudges prevent informed decisions. In their experiments, users were either faced with banners visually promoting a privacy-diminishing option or a reverted interface where the option protecting users' privacy was promoted instead. Related efforts of this community highlight current shortcomings of the GDPR~\cite{eu_gdpr_2016} to achieve its goals. Reviewing compliance of consent management platforms, Nouwens et al.~\cite{nouwens_dark_2020} show that only 11.6\% of websites from a corpus of 10k met the minimum requirements of European law. Reviewing the GDPR for its objectives to give users control over their data, Boyens et al.~\cite{bowyer_human-gdpr_2022} find that users experience serious problems, leading to decreasing trust in institutions that should protect them.

These works collectively show that the responsibility to avoid dark patterns can and should not solely fall onto users. Additional protection needs to come from other sources, such as the better implementation of regulations, while research needs to foster our understanding of dark patterns' origins as well as exploited strategies. We contribute to the latter by turning towards SNSs. Unlike prior work, our study utilises Mathur et al.'s dark pattern characteristics as a framework to learn about users' ability to recognise dark patterns in this domain.

\subsection{Regulating Dark Patterns}
The advantages of interdisciplinary efforts between HCI and legal scholars have recently been shown in Gray et al.'s~\cite{Gray2021} work studying consent banners from multiple perspectives. The negative effects of dark patterns in online contexts are not a new phenomenon in law. Protecting users and consumers from manipulation, unfair practices, and imbalances has always been a subject of legislation. Different laws can affect single design patterns, including data protection law, consumer law, and competition law, depending on their impact on consumers, traders, and personal data~\cite{eu-2022-behaviour,leiser_dark_2021}. Recently, attempts to regulate dark patterns as a whole have arisen. Especially the European Union started to draft legislation that specifically targets dark patterns. The EU's Digital Service Act~\cite{eu_commission_2022_dsa} (DSA) and its proposal for the Data Act~\cite{eu_commission_2022_da} explicitly provide a definition for dark patterns in their recitals.

A key challenge is to legislate patterns that are rapidly evolving while adopting new strategies to pass regulation, yet maintaining their malice. In the context of SNSs, our study draws attention to tools of HCI that could support legal decisions. Picking up on these works, legislators and regulators could utilise the existing knowledge about dark patterns to extend current approaches to protecting peoples' privacy on further problematic designs that potentially harm their well-being. In the presented work, we explore a novel approach to evaluate the malice of interfaces of four SNSs based on high-level characteristics proposed by Mathur et al.~\cite{mathur2021}.

\section{Study 1: Cognitive Walkthrough}
The purpose of this study is to see whether definitions of dark patterns can be used to recognise similar design strategies in domains other than the ones they were initially identified in.
We, therefore, considered four SNSs (Facebook, Instagram, TikTok, and Twitter) where we had six HCI researchers review mobile applications in the form of cognitive walkthroughs~\cite{jaspers2004}. Each researcher was asked to complete ten tasks designed for identifying and recording any instances of dark patterns on the SNSs' mobile applications. The decision to investigate exactly these four SNSs is based on their overall popularity ~\cite{statista_SNS_pop_2021}, comparable features, and similar user bases.  As the experiment was conducted during the COVID-19 pandemic, participants completed their walkthroughs without supervision. Study 1 aims to answer the following research question: Can dark patterns taxonomies be used by experts to identify and recognise instances in SNSs?

\subsection{Reviewers}
For this experiment, we recruited reviewers who have strong expertise in HCI and UX research and design. In a similar fashion to regulators who have to decide whether a problematic interface requires legal action or not, our participants needed to meet the necessary qualifications to identify dark patterns. Their knowledge of best practices in interface design and user experience makes them more susceptible to recognising potential issues compared to users without access to this particular expertise, as shown in prior research~\cite{digeronimo2020, BongardBlanchy2021}. Recruitment involved reaching out to researchers with backgrounds in cognitive science, computer science, and media science who also specialised in HCI research. Participation was on a voluntary basis. In total, we selected six participants (3 female, 3 male) from the authors' professional network. The average age of the panel was 28.33 years ($SD = 1.63$), with an average experience in HCI research of 3.83 years ($SD = 1.47$). All participants worked in academia in HCI-related research labs. Five are of German nationality, while one reviewer is Russian. While all participants had experience in interface design, except for one, none had prior knowledge of dark pattern academic research. Before conducting the study, each participant was provided with the necessary information on the topic before we obtained their consent. To protect them from the unethical consequences of dark patterns, we provided each participant with devices, new accounts for the SNSs, and data to be used during the study. This is further elaborated in subsection 3.2 Preparation.


\subsection{Preparation} 
After receiving their consent for participating in this study, each reviewer received two smartphone devices, a factory reset iPhone X (iOS 14.5) and a Google Pixel 2 (Android 11), with the social media applications already installed to ensure the same version\footnote{Installed versions consistent throughout Study 1: Facebook (iOS: 321.0.0.53.119; Android: 321.0.0.37.119); Instagram: (iOS: 191.0.0.25.122; Android: 191.1.0.41.124); TikTok (iOS:19.3.0; Android: 19.3.4); Twitter (iOS: 8.69.2; Android: 8.95.0-release.00).} was used by each participant. Both iOS and Android devices were used to distinguish between problematic interface designs caused by the applications and those linked to the operating systems. Also, each participant was provided with a new email account and phone number so they could create new user profiles for their assigned platforms. This was done to respect participants' privacy and to avoid customisation of accounts from previous usages that may impact participants' experience and, subsequently, their findings. Lastly, we stored some amount of media content on each device as part of the cognitive walkthrough, affording the participants to create and post content. Again, this ensured that participants did not have to share any personal information with the SNS.

\subsection{Procedure}
One key element of this study is an extracted dark pattern taxonomy based on Mathur et al.'s~\cite{mathur2021} work, including a review of the dark pattern landscape. The taxonomy, featuring 69 distinct types (see Table~\ref{tab:dp_taxonomy}), was given to each reviewer after a one-hour-long introduction to the topic, followed by another hour to resolve unanswered questions mitigating inconsistencies in reviewers' expertise. 
Despite reviewers' backgrounds in HCI-related fields, this introductory session ensured a common understanding of current conceptualisations of dark patterns. After the introduction, each reviewer was handed informational material containing the presented information and the definitions of the 69 dark pattern types. This material is provided in the supplementary material of this paper. To maintain further consistency throughout the study, we created ten tasks reviewers were asked to complete during the cognitive walkthroughs~\cite{jaspers2004}. Five of these tasks were adapted from research conducted by Di Geronimo et al.~\cite{digeronimo2020} that evaluated popular applications on the Google Play Store. Inspired by elements of their methodology, we increased the amount of time each SNS should be investigated to approximately 30 minutes based on a pre-study. This decision allows us to understand the interfaces of the four SNSs on a deeper level. Lastly, each reviewer was assigned two of the four SNSs ensuring that each application was reviewed three times by independent people on both iOS and Android operating systems. After a reviewer completed their walkthrough, we saved the stored recording data from the devices before setting them up for the next session. Below are the ten tasks each reviewer performed. Tasks taken from or worded closely to Di Geronimo et al.~\cite{digeronimo2020} are highlighted by an asterisk. Items 1, 9, and 10 were added to improve the task flow, whilst items 4 and 5 were developed to address typical SNS activities such as creating and sharing personal content and networking.

\begin{em}
\begin{enumerate}
    \item[1.] Turn on screen recording on each device.
    \item[*2.] Open the app and create an account to log in and then out.
    \item[*3.] Close and reopen the app.
    \item[4.] Create any kind of content, post it, and delete it.
    \item[5.] Follow and unfollow other accounts.
    \item[*6.] Visit the personal settings.
    \item[*7.] Visit the ad-related settings.
    \item[*8.] Use the application for its intended use (minimum of five minutes):
    \begin{enumerate}
        \item[I] Describe the natural flow of the app – what did you use it for?
        \item[II] Could you use the app as you wanted or did some features 'guide' your interactions?
        \item[III] how easy was it to get distracted and if so what distracted you?
    \end{enumerate}
    \item[9.] Delete your account.
    \item[10.] Turn off screen recording and save the recording.
\end{enumerate}
\end{em}

\section{Results of Study 1}

\begin{figure*}[t!]
    \centering
    \subfloat[\centering Screenshot from Facebook]
    {{\includegraphics[width=0.22\textwidth]{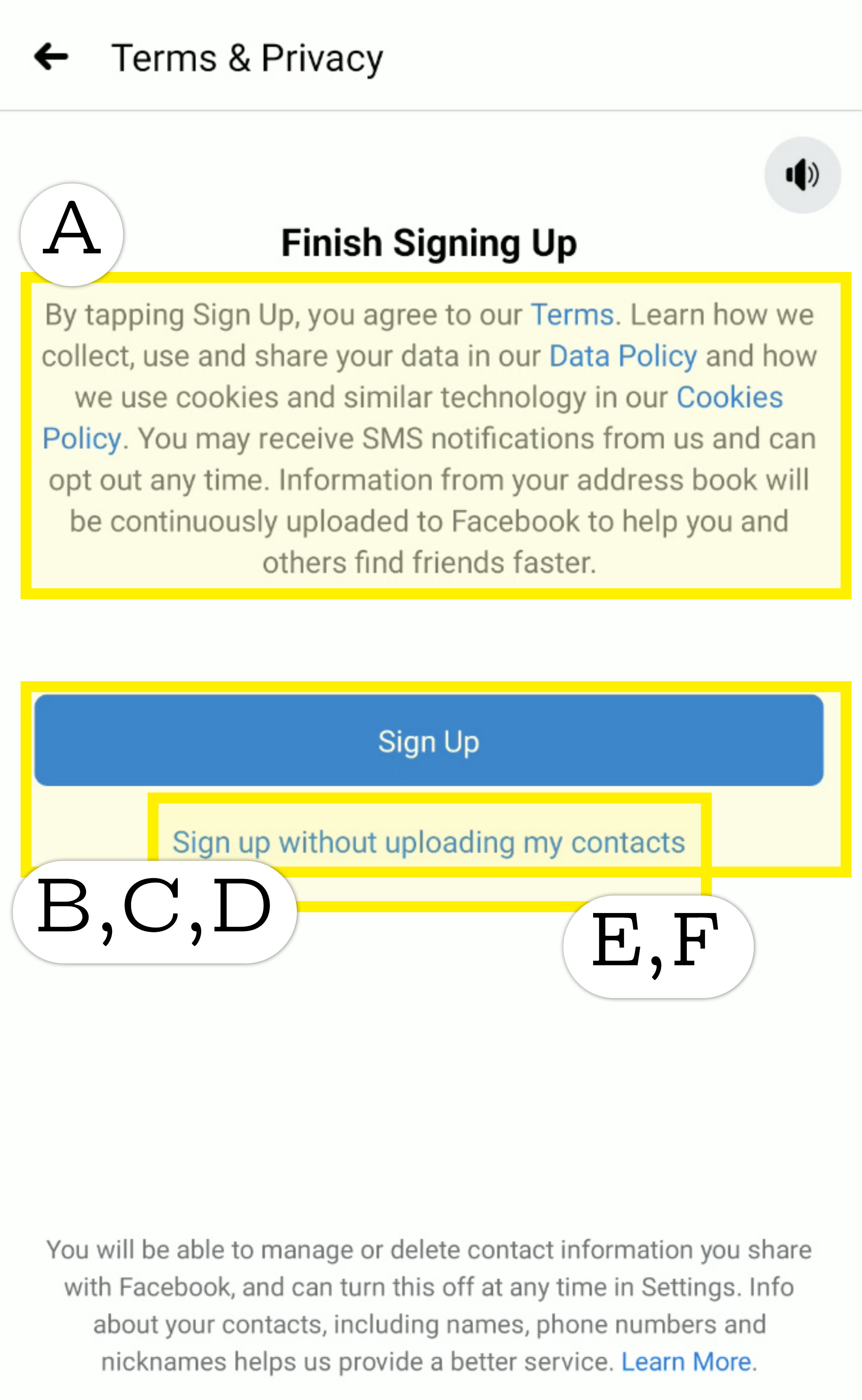}\label{fig:ex-facebook}}}\hfill%
    \subfloat[\centering Screenshot from Instagram]
    {{\includegraphics[width=0.22\textwidth]{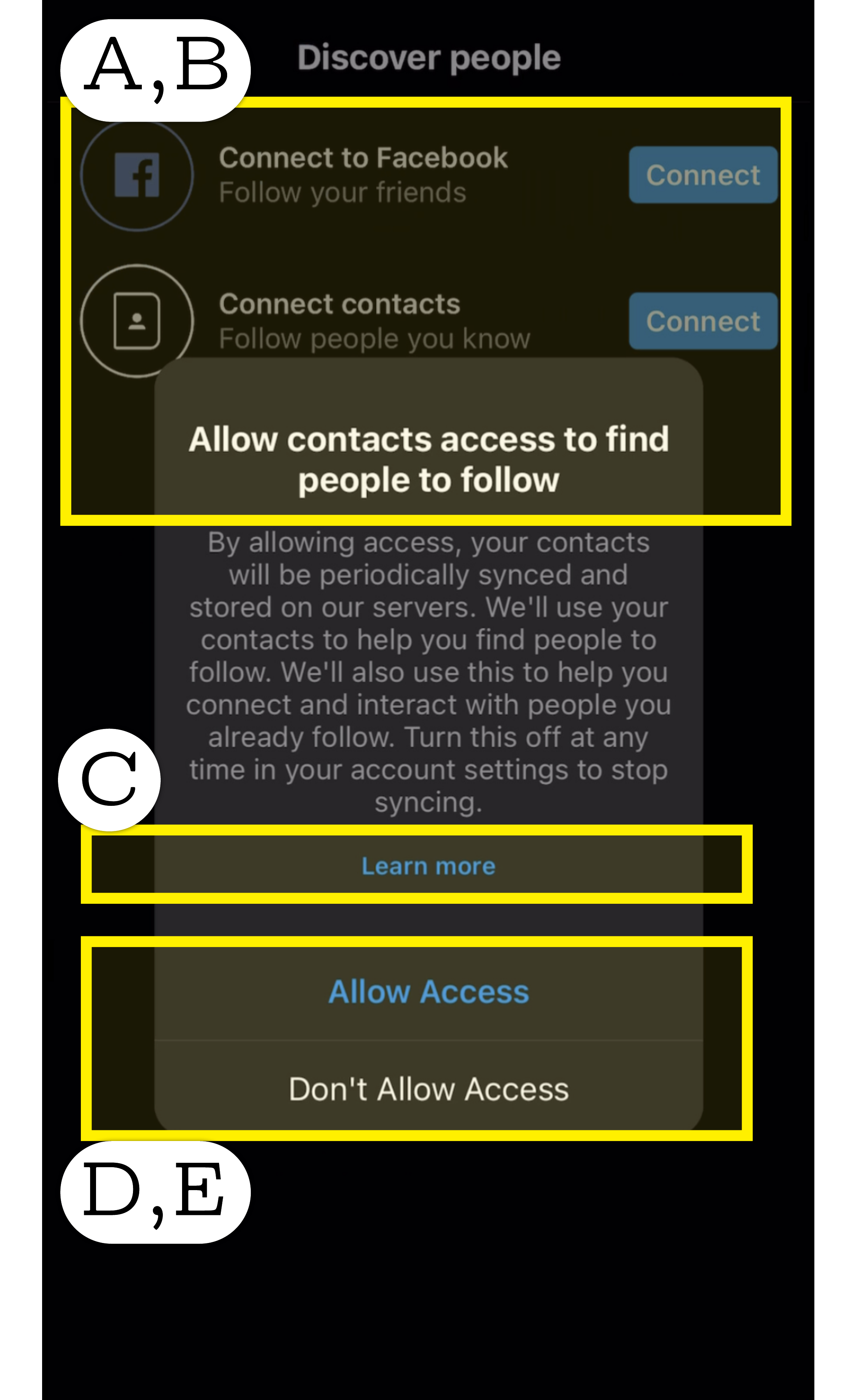}\label{fig:ex-instagram}}}\hfill%
    \subfloat[\centering Screenshot from TikTok]
    {{\includegraphics[width=0.22\textwidth]{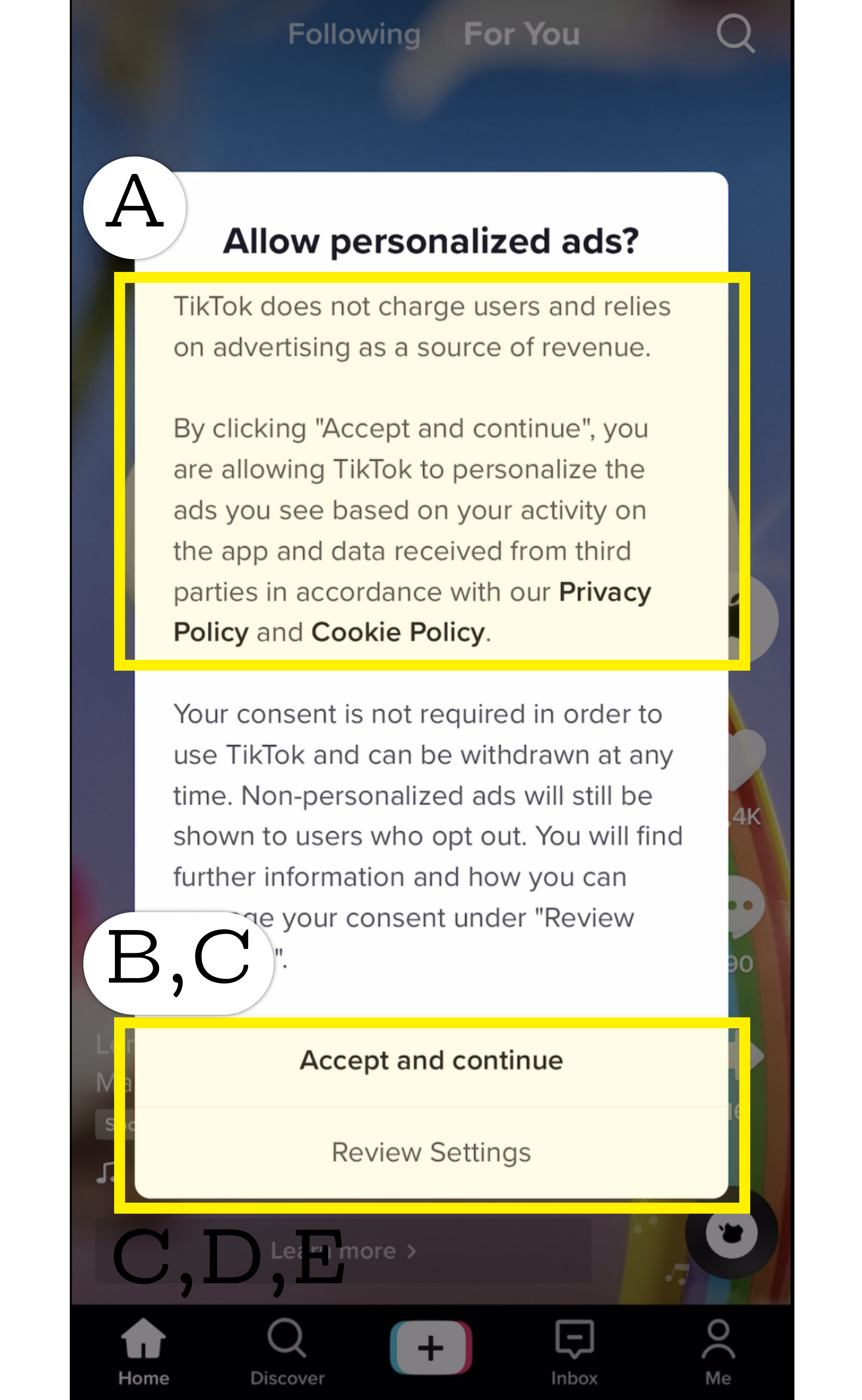}\label{fig:ex-tiktok}}}\hfill%
    \subfloat[\centering Screenshot from Twitter]
    {{\includegraphics[width=0.22\textwidth]{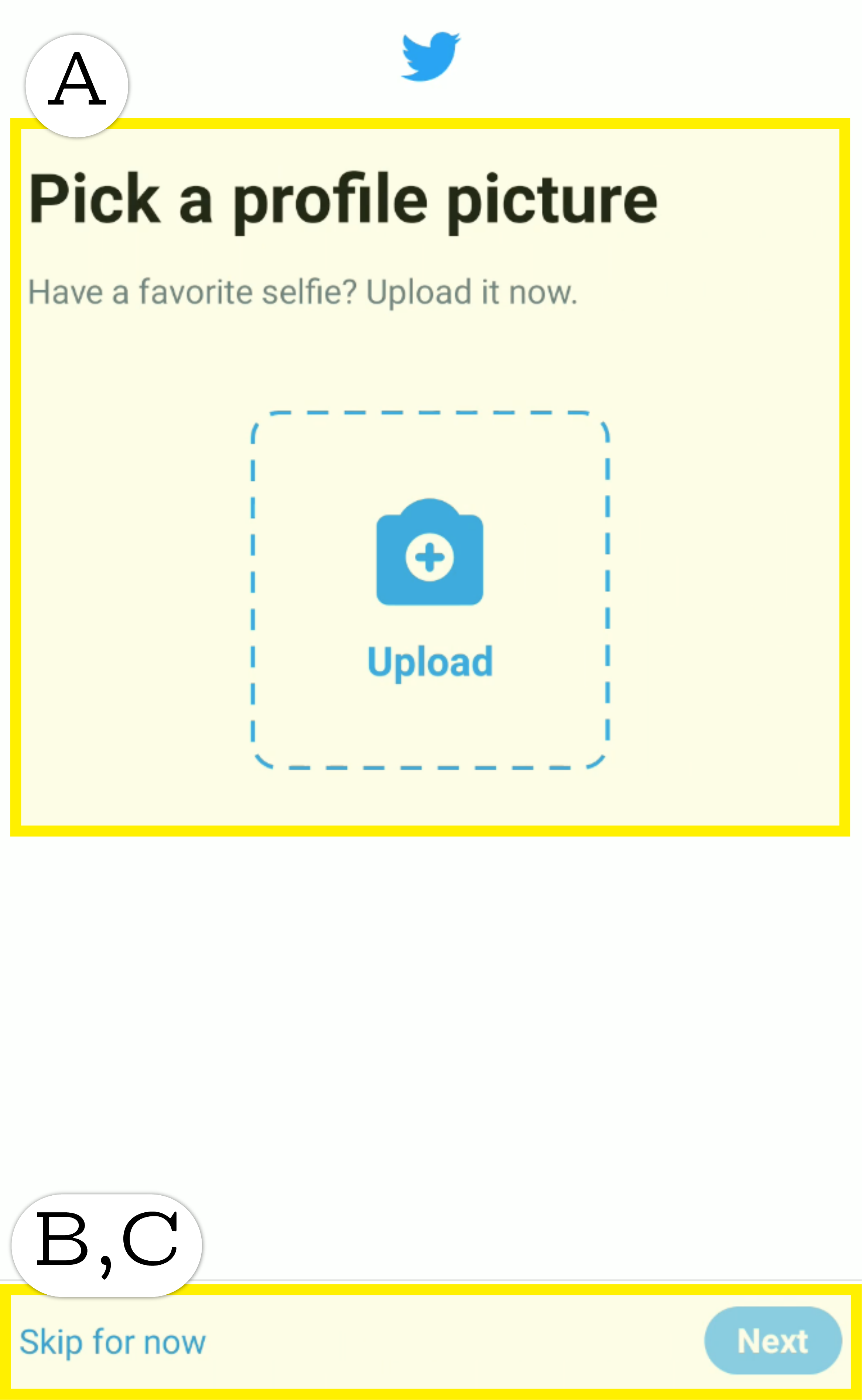}\label{fig:ex-twitter}}}\hfill%
    \caption{Example screenshots from Study 1. 
    \textbf{Figure \ref{fig:ex-facebook}} contains the dark patterns \textit{Hidden-Legalese Stipulations (A), Misdirection (B), Interface Interference (C), Visual Interference (D), Privacy Zuckering (E),} and \textit{Address Book Leeching (F)}. 
    \textbf{Figure \ref{fig:ex-instagram}} contains the dark patterns \textit{Privacy Zuckering (A), Address Book Leeching (B), Hidden-Legalese Stipulation (C),  Interface Interference (D), and Visual Interference (E)}.
    \textbf{Figure \ref{fig:ex-tiktok}} contains the dark patterns \textit{Hidden-Legalese Stipulation (A), Interface Interference (B)} and \textit{Visual Interference (C)}.
    \textbf{Figure \ref{fig:ex-twitter}} \textit{Privacy Zuckering (A), Interface Interference (B), and Visual Interference (C)}.
    }
    \Description[Screenshots of the four mobile apps with dark patterns.]{This Figure shows four example screenshots from Study 1 containing dark patterns. For each screenshot, opaque yellow boxes highlight where dark patterns can be seen. Alphabetic letters enumerate the dark patterns. Figure 2a shows a prompting interface from Facebook asking the user to finish setting up their account. It contains a long text in light grey containing links to legal information. Here, the links are highlighted in blue. It further has a prominent blue button saying sign up. Below is a second option, although not within a big blue box and thus somewhat obfuscated, saying 'Sign up without uploading my contacts', implying that the big button would automatically upload the users' contacts. This interface contains the dark patterns Address Book Leeching, Hidden-Legalese Stipulations, Misdirection, Privacy Zuckering, Interface Interference, and Visual Interference. Figure 2b shows an interface from Instagram prompting the user to allow the app to access the users' local contacts to find and suggest people. A light grey includes more legalese information, although no specific links are provided. Instead, a 'lean more' link suggests that information may be found elsewhere. At the bottom of the interface, the user is given two options. In more prominent blue, they can allow access; in light grey, they can decline. This interface contains the dark patterns Confirmshaming, Interface Interference, and Visual Interference. Figure 2c visualises a screenshot from TikTok. The interface asks the user to allow personalised ads. Similar to previous interfaces, it begins with a light grey information text box in the centre that includes links to legal information like privacy and cookie policies. At the bottom, it presents the user with two options: They can either accept or continue, meaning they accept personalised ads. This option is light blue. The alternative says review settings implying the user can make additional choices elsewhere. However, this option is obscured by light grey colouring. The interface contains dark patterns Interface Interference and Visual Interference. Figure 2d shows a screenshot from Twitter. It prominently asks the user to pick a profile picture, including a big button to upload a photo directly. This part of the interface takes about 60 per cent of the screen's space. At the bottom, the user can move forward by clicking a blue next button in the bottom-right corner. At the bottom-left corner, a less obvious option to skip this request is provided. This screenshot contains the dark patterns Privacy Zuckering, Interface Interference, and Visual Interference.}
    \label{fig:ex-screenshots}
\end{figure*}

In this study, we considered a dark pattern taxonomy comprising 69 individual types of dark patterns (see Table~\ref{tab:dp_taxonomy}) across mobile applications for the SNSs Facebook, Instagram, TikTok, and Twitter. Offering an answer to our first research question, the six participants identified a total of $548$ dark pattern distinct instances from the considered 69 types that can be associated with descriptions contained within the taxonomy provided. Participants found $N_{F}=232$ dark pattern instances in Facebook, $N_{I}=96$ in Instagram, $N_{Ti}=95$ in Twitter, and $N_{Tw}=125$ in Twitter. Figure~\ref{fig:ex-screenshots} presents four screenshots that demonstrate examples of dark patterns identified by participants across each of the four SNSs. Close inspection shows multiple types of dark patterns at play in each image. Although the four SNSs were selected based on similar functionalities and user bases, we do not compare results across platforms. Despite their similarities, each SNS contains unique features that distinguishes them from the others. Also, the number of functionalities between the SNSs varies considerably, with Facebook containing many more options for users to engage with than alternatives. Instead,
we report descriptive statistics that will then be further elaborated on in the discussion section of this paper.

\subsection{Recognised Types of Dark Patterns}

\begin{figure}[]
    \centering
    \includegraphics[height=0.915\textheight]{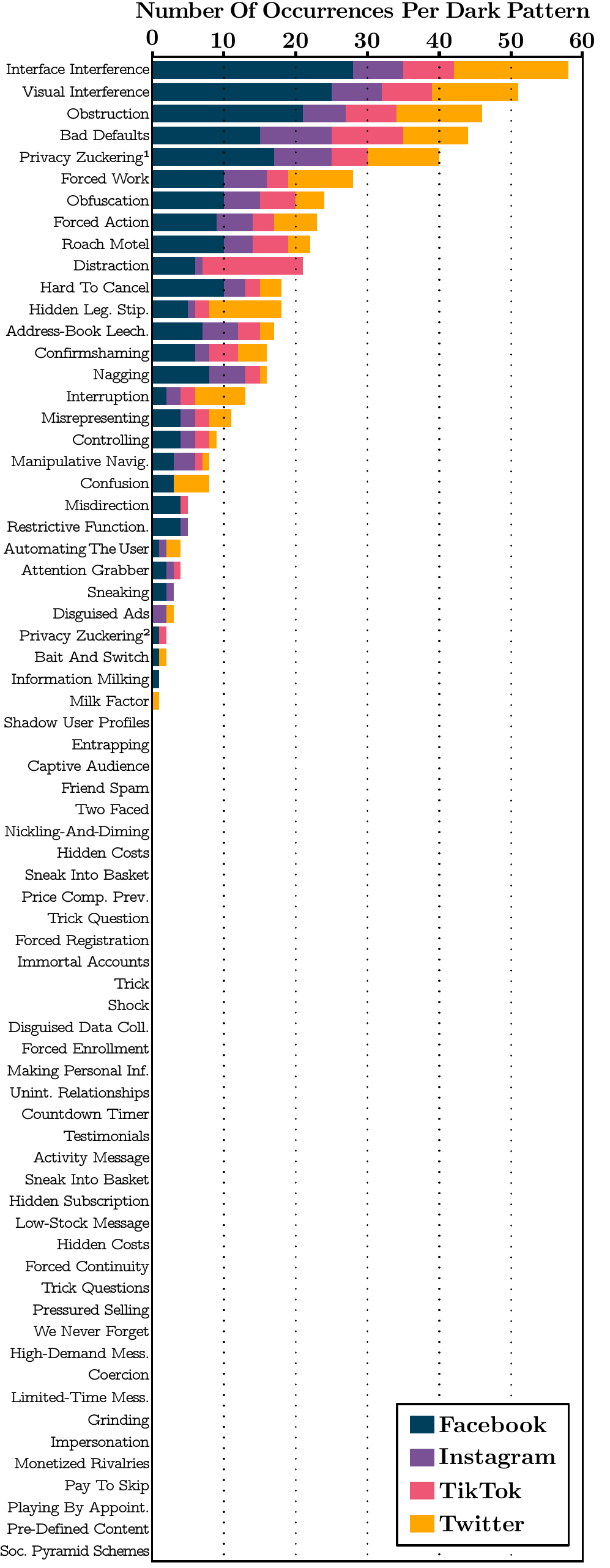}
    \caption{Summary of the occurrences of all 69 considered dark pattern types in four SNSs.
    Of the 69 types 31 were recognised. \textit{Privacy Zuckering\textsuperscript{1}} refers to Brignull's~\cite{brignull_deceptive_nodate} description while  \textit{Privacy Zuckering\textsuperscript{2}} refers to Bösch et al.'s defintion~\cite{bosch2016}.}
    \label{fig:occurrence_of_dp_in_sns}
    \Description[Occurrences of dark patterns in SNSs.]{This Figure shows a vertical, stacked bar chart representing the total occurrences of dark patterns in SNSs (Facebook in dark blue, Instagram in purple, TikTok in magenta, and Twitter in yellow) based on the considered taxonomy of 69 dark patterns. Foremost, the chart demonstrates which 31 dark patterns were recognised in the SNSs and which 31 did not.}
\end{figure}

Of the 69 types of dark patterns contained in the taxonomy participants were provided with at the beginning of this study, 31 distinct types were identified, leaving the remaining  $55.07\%$ unrecognised across any of the four SNSs. All recognised dark patterns can be seen in Figure~\ref{fig:occurrence_of_dp_in_sns}. 
For brevity, only key illustrative instances are reported here, while the full analysis will be included in the supplementary material. Across the four SNSs, two dark pattern types stood out the most: With a total of 58 recognised instances, Gray et al.'s \textit{Interface Interference}~\cite{gray2018} (i.e. interfaces that privilege certain elements over others confusing users to make a particular choice) was most readily identified by participants, whilst Mathur et al.'s \textit{Visual Interference}~\cite{mathur2019} (i.e. interfaces that deploy visual/graphical tricks to influence users' choices) was next most widely observed with 51 instances. The third most frequently identified dark pattern was Gray et al.'s \textit{Obstruction}~\cite{gray2018} dark pattern (interfaces that make certain actions unnecessarily difficult to demotivate users) recognised 47 times. Bösch et al.'s \textit{Bad Defaults}~\cite{bosch2016} (privacy settings are pre-set to share users' personal information by default) came fourth with 44 instances, closely followed by 40 counts of Brignull's \textit{Privacy Zuckering}~\cite{brignull_deceptive_nodate} (tricks to deceive users into sharing more personal information than intended) dark pattern.


\subsection{Types of Dark Patterns That Have Not Been Recognised}

 
While 44.93\% of dark pattern types were recognised during the cognitive walkthrough, the other  55.07\% were not. 
Almost all dark pattern taxonomies contained some dark patterns that were recognised. However, the taxonomy by Zagal et al.~\cite{zagal_dark_2013}, being video-game focused, did not contribute any specific dark patterns that were recognised. This result shows that not all dark pattern types are relevant for each domain. By adding new dark pattern types to the overall collection for each domain, regulators have increasingly more items to consider complicating their endeavour if they are to use them as guides.

\subsection{Dark Patterns Co-Occurrences}
To learn more about how dark patterns interact with each other, we also analysed them for co-occurrences. We used the software ATLAS.ti~\cite{atlasti_2021} to calculate the co-occurrence coefficient between any two dark patterns, which is based on the Jaccard similarity coefficient~\cite{friese_2019} returning a c-coefficient $c$. Interestingly, the data revealed that although two patterns are described differently, their working can be rather similar in the context of SNSs. Intersections between $\textit{Interface Interference} \cap \textit{Visual Interference}$ ($c=0.85$, $N=50$ co-occurrences), \textit{Forced Action} $\cap$ \textit{Forced Work} ($c=0.89$, $N = 25$ co-occurrences), and \textit{Roach Motel} $\cap$ \textit{Hard to Cancel} ($c=0.71$, $N=17$ co-occurrences), for instance, follow this example. However, like the intersection between \textit{Misrepresenting} $\cap$ \textit{Immortal Accounts} ($c=0.55$, $N=12$ co-occurrences) or \textit{Privacy Zuckering} $\cap$ \textit{Bad Defaults} ($c=0.35$, $N=22$ co-occurrences), most co-occurrences are indications for interfaces yielding multiple distinct dark patterns simultaneously. Due to the overall co-occurrence data set is too large to be fully represented here, it has been included in the supplementary material.


\section{Study 2: Online Survey}
Findings from Study 1 suggest existing taxonomies feature numerous types of dark patterns that are not applicable to SNSs and that some dark patterns employed by SNSs are not incorporated in earlier taxonomies. In this second study, we adopted a different approach to identifying dark patterns in interfaces. Instead of relying on fixed descriptions and definitions of existing dark patterns, we developed a questionnaire consisting of five questions based on dark pattern characteristics previously highlighted by Mathur et al.~\cite{mathur2021}. These higher-level characteristics go beyond dark pattern definitions by descriptively organising dark patterns from existing literature~\cite{mathur2021}. Following this approach, study 2 aims to address the following research question: Are regular SNS users able to differentiate between interfaces with and without dark patterns?

\subsection{Screenshots}

\begin{figure*}[t!]
    \centering
    \subfloat[\centering Screenshot With Dark Patterns - Facebook]
    {{\includegraphics[width=0.22\textwidth]{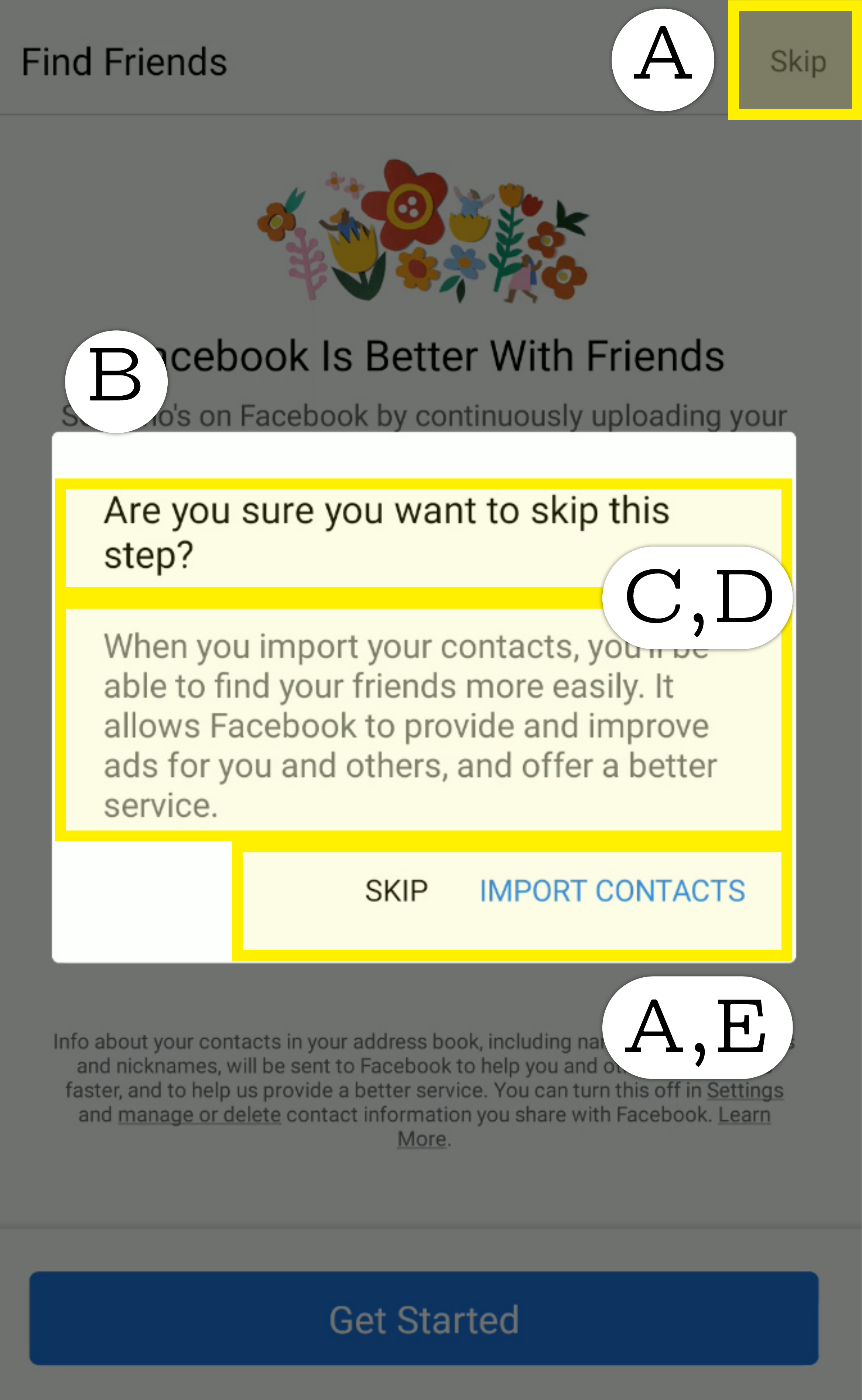}\label{fig:dp-facebook2}}}\hspace{10pt}%
    \subfloat[\centering Screenshot With Dark Patterns - Twitter]
    {{\includegraphics[width=0.22\textwidth]{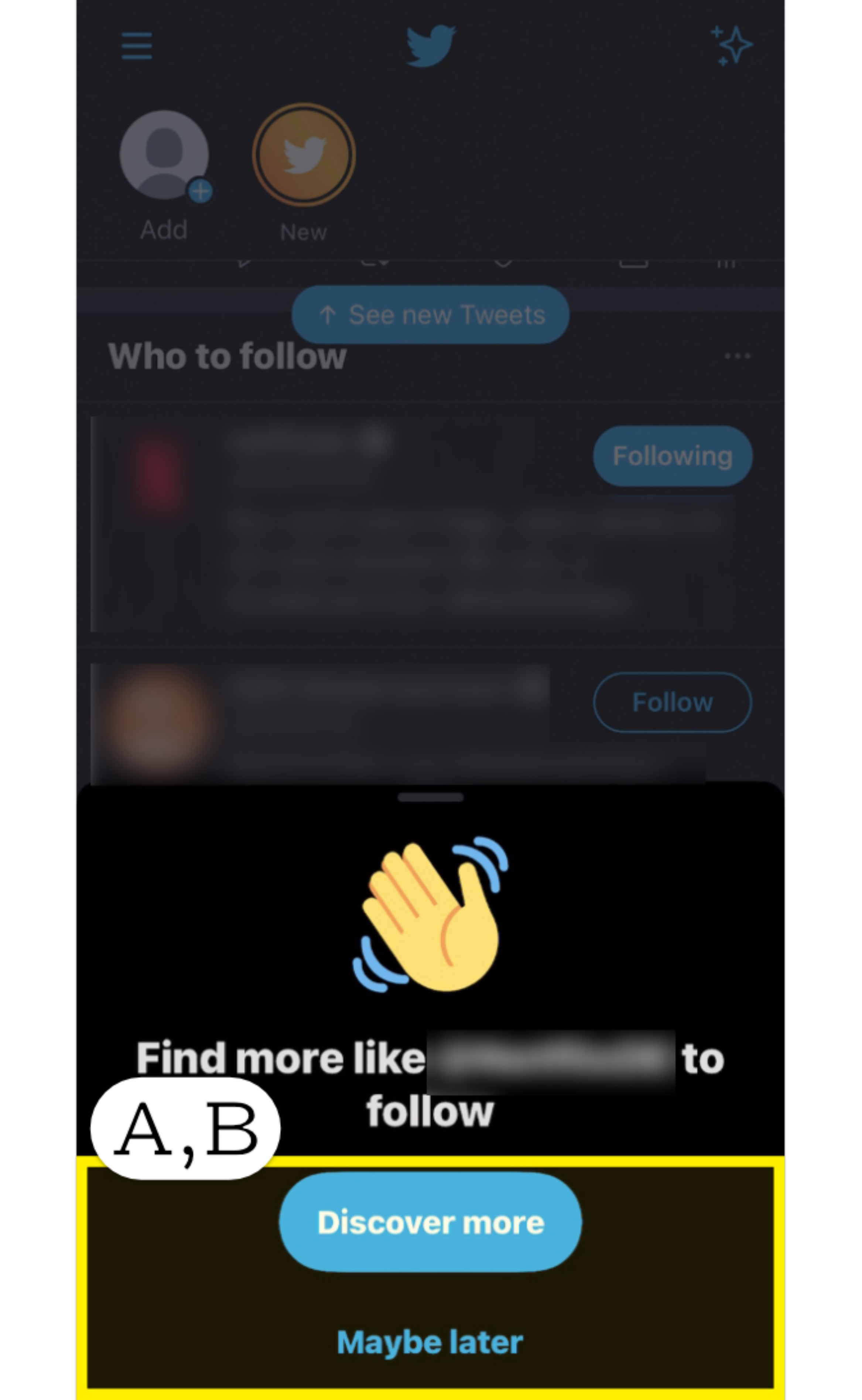}\label{fig:dp-twitter2}}}\hfill%
    \subfloat[\centering Screenshot Without Dark Patterns - Facebook]
    {{\includegraphics[width=0.22\textwidth]{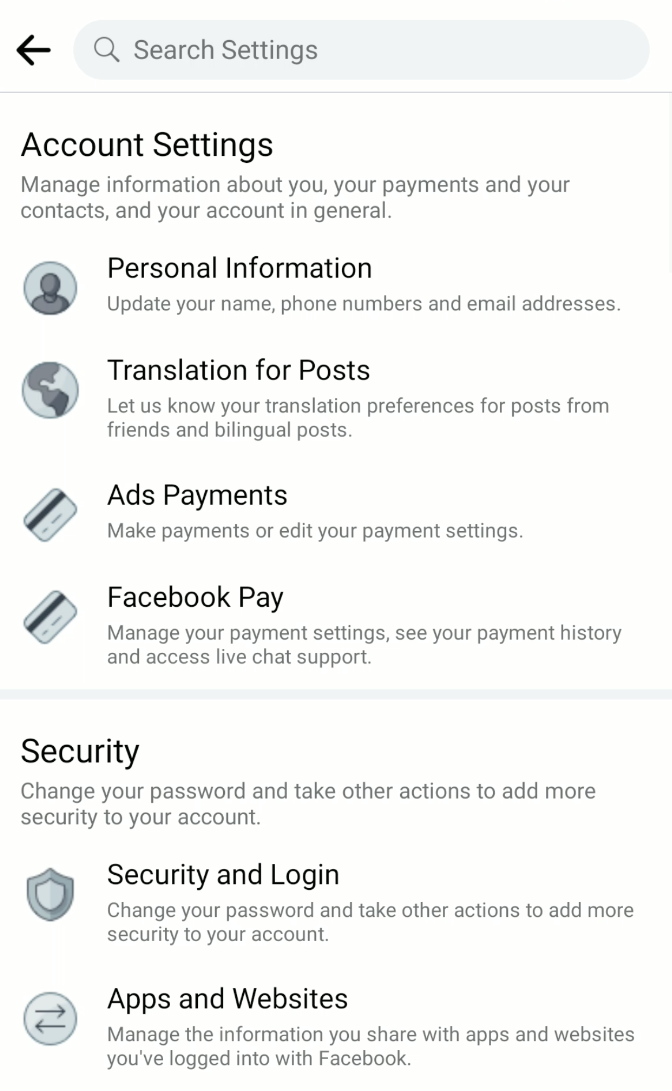}\label{fig:np-facebook}}}\hspace{10pt}%
    \subfloat[\centering Screenshot Without Dark Patterns - Instagram]
    {{\includegraphics[width=0.22\textwidth]{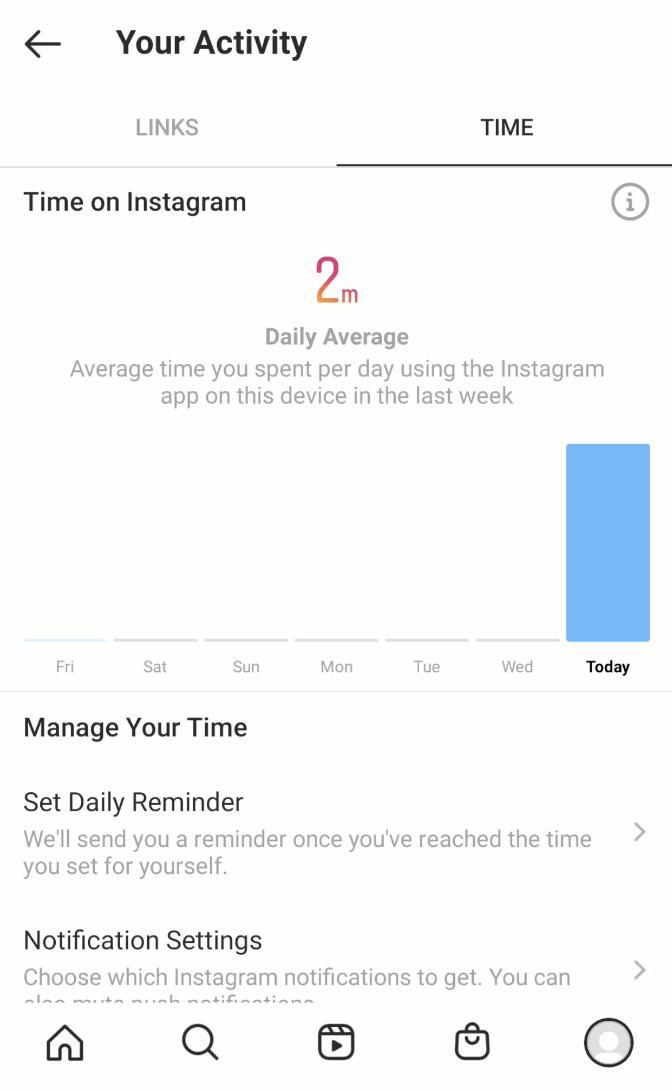}\label{fig:np-instagram}}}%
    \caption{Four example screenshots used in study 2, sampled from study 1.
    \textbf{Figure \ref{fig:dp-facebook2}} contains the dark patterns \textit{Interface Interference (A), Confirmshaming (B), Address-Book Leeching (C), Privacy Zuckering (D),} and\textit{Visual Interference (E)}. 
    \textbf{Figure \ref{fig:dp-twitter2}} contains the dark patterns \textit{Interface Interference (A)}, and \textit{Visual Interference (B)}. Importantly, \textbf{Figure \ref{fig:dp-facebook2}} and \textbf{Figure \ref{fig:dp-twitter2}} were presented to participants without annotations.
    Neither \textbf{Figure \ref{fig:np-facebook}} nor \textbf{Figure \ref{fig:np-instagram}} contain any dark patterns. In total, sixteen screenshots were used in study 2 - eight containing dark patterns and eight that do not.
    }
    \Description[Example screenshots used in study 2, sampled from study 1.]{
    Figure 3a is a screenshot taken from Facebook. It visualises a scenario where a participant denied upload of their contacts resulting in a prompt asking them to confirm this decision. The question reads: Are you sure you want to skip this step? However, options are obscured through visually highlights of an import contacts button in bright blue over a skip button in light grey. The interface contains the dark patterns Interface Interference, Confirmshaming, Address-Book Leeching, Privacy Zuckering, and Visual Interference. Dark patterns are highlighted with transparent yellow boxes and given alphabetic labels to describe them in the Figure's caption. Figure 3b contains shows a screenshot from Twitter. It demonstrates the dark patterns Interface Interference and Visual Interference. The image is an impromptu suggestion to the SNSs to find more accounts to follow. Options are again visually displayed differently. The discover more option is a big circular blue button, whereas the alternative option (maybe later) is without any visual aids and only represented by its text. Neither Figure 3c nor Figure 3d contains any dark patterns. Importantly, Figure 3a nor 3b were presented to participants without annototations. Figure 3c shows the settings menu of Facebook featuring links to account and security-related settings. Figure 3d stems from Instagram and shows the usage activity of a participant with regards to the amount of time they have been using the application. A bar chart represents the time for each day of the past week. Also, links to settings for daily reminders and notifications are contained.}
    \label{fig:comparison-screenshots}
\end{figure*}

We used sixteen screenshots along with the aforementioned questionnaire to evaluate people's ability to recognise dark patterns within screenshots of the four SNSs. While eight of the sixteen screenshots contained dark patterns, the other eight did not and served as control. All screenshots were sampled from the previous study (see Figure~\ref{fig:comparison-screenshots} for four example images). Regarding those that contained dark patterns, two conditions had to be met: Screenshots had to (1) represent all five characteristics by Mathur et al. while (2) contained dark patterns had to be identified by at least two expert reviewers. Furthermore, we avoided using screenshots that contained dark patterns that only emerge through procedural interactions taken by users (e.g. \textit{Roach Motel}). Consequently, two authors of this paper ensured to pick screenshots where the dark patterns were recognisable on a static image, for example by deploying visual/aesthetic (e.g.Visual Interference) or linguistic (e.g. Confirmshaming) manipulations.
Screenshots that did not contain dark patterns were carefully selected by sampling situations where expert reviewers did not recognise any dark pattern. This was additionally validated by two authors of this paper to ensure no dark pattern had been accidentally overlooked. Using these screenshots, we test whether participants can generally recognise dark patterns and  whether they can differentiate between screenshots with and without dark patterns. 

\subsection{Methodology}
To investigate our research question, we conducted an online survey. The survey was divided into three parts: (1) screening for participants' SNS usage behaviour, (2) a dark pattern recognition task, and (3) a demographic questionnaire. In total, the survey featured 25 question items (included in supplementary material) and took on average 12:22 minutes ($SD=$ 9:45) to complete. As we were interested if regular social media users could asses dark patterns in SNS, only participants who indicated previous and regular use of social media platforms were included in the sample. This was achieved using screening questions about previous social media usage. Before evaluating the sixteen screenshots, participants were provided with the following definition of dark patterns by Mathur et al.'s~\cite{mathur2021}: \textit{``user interface design choices that benefit an online service by coercing, steering, or deceiving users into making decisions that, if fully informed and capable of selecting alternatives, they might not make''}. For each of the sixteen screenshots, participants had to first answer if they thought dark patterns were present in the screenshot based on the definition of dark patterns by Mathur et al.'s~\cite{mathur2021} with 'Yes', 'No' or 'Maybe'. In the next step, participants then had to answer if they saw dark patterns in the screenshot based on Mathur's dark pattern characteristics~\cite{mathur2021}. For this, we developed five questions adopting the characteristics~\cite{mathur2021}, which participants rated based on a unipolar 5-point Likert-scale (see Table~\ref{tab:mathur2019}). Available responses ranged from ``Not at all'' to ``Extremely''. After assessing all five characteristics, they moved on to the next screenshot. Screenshots were delivered in a randomised order between participants. Once all screenshots were assessed, the survey concluded by collecting basic demographic data from each respondent, including age, gender, current country of residency, and an optional field to give feedback.

\subsection{Participants}
To calculate an appropriate sample size needed to answer our research questions, we conducted an \textit{a priori} power analysis using the software G*Power ~\cite{gpower_2007}. Given our study design, to achieve a power of 0.8 and a medium effect size, the analysis suggested a total sample size of 166. Participants of this survey were recruited from two sources: (1) The Reddit forum \textit{r/samplesize}~\cite{reddit_samplesize} and (2) \textit{Prolific}~\cite{prolific}. For redundancy, we invited 90 people, more than our power analysis suggested. After receiving their consent to participate in this study, 256 participants were recruited and completed the online survey. Of these 256 participants, 26 were recruited via Reddit~\cite{reddit_samplesize} and 230 via Prolific~\cite{prolific}. Initially, we recruited participants from Reddit to assess the feasibility of our study design. After this was ensured and we successfully verified that the retrieved data was equal in quality to the data gained from Prolific, both sets were accumulated. Compensation for participating in this study was rewarded with \pounds{7.2} per hour, with individual compensation dependent on participants' time needed to complete the study (mean $= 12.2$ minutes, $SD = 8.76$ minutes). We excluded 63 data sets in total due to: failure to complete the questionnaire; failed attention checks (questions with a single true answer to measure participants' engagement); not meeting inclusion criteria; completing the questionnaire in unrealistic times based on \textit{a priori} testing; and if they replied with the same option over $95\%$ of instances. Eventually, data from a total of 193 participants were included in the analysis, thus satisfying the estimate of the power analysis.

\section{Results of Study 2}

\begin{table}[t]
\renewcommand{\arraystretch}{1.4}
\begin{tabular}{p{0.2\linewidth}p{0.7\linewidth}}
\toprule
\multicolumn{2}{c}{{\begin{tabular}[c]{@{}c@{}}\textbf{Mathur 2019 \cite{mathur2019}}\\ Dark Pattern Characteristics\end{tabular}}}                                \\ \midrule
Characteristic  & Question                                                                                                                          \\ \midrule
\cellcolor[gray]{0.95}Asymmetric          & \cellcolor[gray]{0.95}Does the user interface design impose unequal weights or burdens on the available choices presented to the user in the interface? \\
Covert              & Is the effect of the user interface design choice hidden from the user? \\
\cellcolor[gray]{0.95}Deceptive           & \cellcolor[gray]{0.95}Does the user interface design induce false beliefs either through affirmative misstatements, misleading statements, or omissions?\\
Hides Information   & Does the user interface obscure or delay the presentation of necessary information to the user? \\
\cellcolor[gray]{0.95}Restrictive         & \cellcolor[gray]{0.95}Does the user interface restrict the set of choices available to users?  \\ 
\bottomrule
\end{tabular}
\caption{This table lists the introductory questions Mathur et al. (2019)~\cite{mathur2019} gave for each dark pattern characteristic.}
\Description[Dark Pattern Characteristics]{This table lists the introductory questions Mathur et al. (2019) gave for each dark pattern characteristic. The characteristics and their introductory questions are as follows: 
1. Asymmetric: Does the user interface design impose unequal weights or burdens on the available choices presented to the user in the interface?

2. Covert: Is the effect of the user interface design choice hidden from the user?

3. Deceptive: Does the user interface restrict the set of choices available to users?

4. Hides Information: Does the user interface design induce false beliefs either through affirmative misstatements,
misleading statements, or omissions?

5. Restrictive: Does the user interface obscure or delay the presentation of necessary information
to the user?}
\label{tab:mathur2019}
\end{table}

In this section, we present the results of the online survey. The results are split into three parts: (1) demographic data on our participants, (2) results on whether participants can recognise dark patterns based on the definition of dark patterns by Mathur et al.~\cite{mathur2021}, and (3) whether they can differentiate between screenshots with and without dark patterns based on Mathur's dark pattern characteristics (see Table \ref{tab:mathur2019}), as a recognition task including the 69 different individual dark pattern types would have exceeded the scope and purpose of this online survey. Instead, we relied on Mathur et al.'s high-level dark pattern characteristics. For each of the five dark pattern characteristics (\textit{asymmetry}; \textit{covert}; \textit{deception}; \textit{information hiding}; and \textit{restriction}) participants rated on a 5-point Likert scale (``Not at all'' - ``Extremely''), how much the characteristic was present in the screenshot. For each screenshot, this resulted in an average rating. Figure~\ref{fig:dp_vectorisation} demonstrates how the screenshots were used to generate these ratings. This procedure allows us to compare participants' ratings between the different screenshots. Using this approach, the maximum rating for a screenshot featuring all dark pattern characteristics corresponds to $[4,4,4,4,4]$ and thus an average rating of $4$, while a minimum rating for a screenshot without dark patterns corresponds to $[0,0,0,0,0]$ and thus an average rating of $0$. In total, all 193 survey respondents rated ($193*16=3088$) $3088$ screenshots.

\subsection{Demographic Information}
The mean age across individuals was $\mu=27.91$ years ($SD=9.53$), with 155 identifying as female and 35 as male. 
The remainder (N=3) identified as either non-binary or with a third gender. When asked about their current country of residence, the participants replied as follows: Australia (4); Canada (35); France (1); Greece (1); Hong Kong - S.A.R. (1); Ireland (11); Japan (1); South Africa (2); Spain (1); United Kingdom of Great Britain and Northern Ireland (40); United States of America (96). In terms of how frequently participants used the internet, 189 self-reported using the internet on a daily basis, with the remainder (N=4) using it more than once per week. An inclusion criterion for participation was a previous experience with at least one of the four SNSs.  Therefore, we asked participants about their usage of Facebook, Instagram, TikTok, and Twitter. Regarding Facebook, 138 participants reported actively using it, 20 do not use it, and 35 used to use it but not anymore. 167 participants currently use Instagram, while 15 do not use it, and 11 have used it but do not anymore. Looking at TikTok, 134 participants use it currently, 55 do not, and 4 have used it but do not anymore. Lastly, 112 participants actively use Twitter, 51 are not using it, whereas 30 used to but do not anymore.

\subsection{Generally Recognising Dark Patterns} \label{sec:results-rec-dp}
\begin{figure}[ht]
    \centering
    \includegraphics[width=0.45\textwidth]{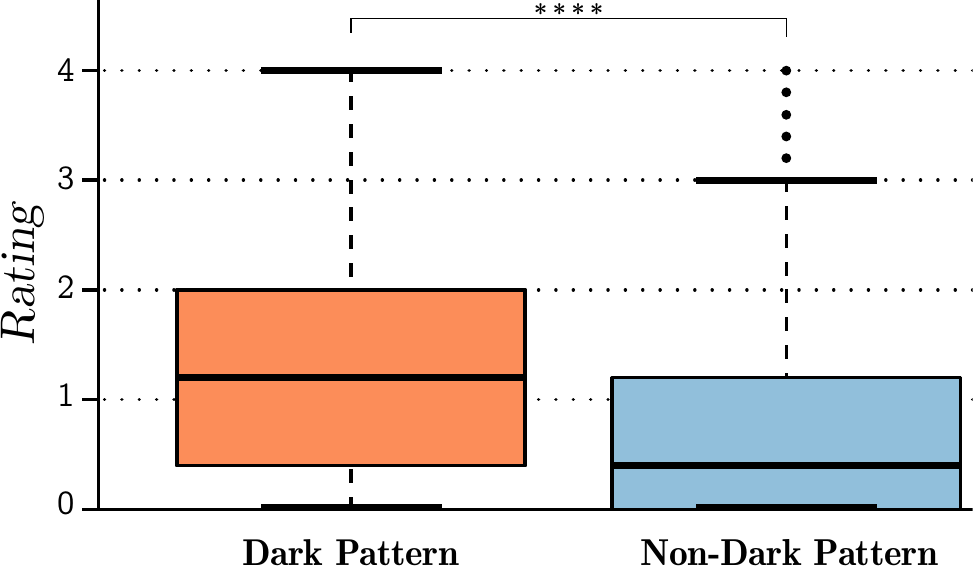}
    \caption{This box plot visualises the differences in which participants, who were provided with a definition for dark patterns, rated the screenshots after being asked if they noticed any malicious designs. The figure shows a significant difference between participants' ratings of screenshots containing dark patterns versus those that do not.}
    \Description[Two box plots showing participants' ability to differentiate dark pattern screenshots from non-dark pattern screenshots.]{This figure shows two box plots that demonstrate participants' ability to recognise dark patterns based on screenshots, given a definition of dark patterns of Mathur et al. (2021). In orange, the first box plot shows participants' median ratings for screenshots containing dark patterns. A blue box plot shows their median rating of screenshots that did not contain dark patterns. There is a significant difference (p < 0.0001) in their ratings, suggesting that they can recognise dark patterns and differentiate between different screenshots.}
    \label{fig:dp-vs-ndp}
\end{figure}

For the eight screenshots that did feature dark patterns, when asked if respondents notice any malicious interface elements in the screenshot, 426 screenshots received a ``yes'' rating, 408 a ``maybe'', and 710 a ``no'' rating. In contrast, for the eight screenshots that did not contain dark patterns, 143 received a ``yes'' rating, 269 a ``maybe'', and 1132 a ``no'' rating. A Wilcoxon signed rank test with continuity correction shows significant differences between the two groups of screenshot ratings ($V = 89253$, $p-value < 0.0001$, $R = 0.37$). Thus, we see that more people noticed malicious elements in screenshots that contained dark patterns.

\subsection{Differentiating Between Screenshots With and Without Dark Patterns}
Our previous results showed that people generally see differences between the two types of screenshots. We can thus test whether people rate screenshots differently when they show dark patterns compared to screenshots with no dark patterns according to Mathur et al.'s \cite{mathur2019} five characteristics.  We thus calculated the median total rating for screenshots that featured dark patterns and the same for screenshots that did not feature dark patterns. Across all screenshots which featured dark patterns, we find a median rating of $1.2$ ($mean = 1.26$, $SD = 1.02$) compared to a median rating of $0.2$ ($mean = 0.69$, $SD = 0.81$) for screenshots without dark patterns (see Figure~\ref{fig:dp-vs-ndp}). A Wilcoxon signed-rank test results in a significant difference between the two ratings ($V = 669900$, p-value $< 0.0001$, $R = 0.3$). Given that non-dark pattern screenshots received a significantly lower median average rating than dark pattern screenshots, we conclude that people recognised a difference between screenshots containing dark patterns and those that did not base on questions adopting the five characteristics. We further observe a difference in participants' perceptions of the two types of screenshots. While the median rating of screenshots without dark patterns is $0.2$, very close to $0$ (``Not at all''), the median rating of screenshots with dark patterns is $1.2$ (``A little bit''), relatively low considering a maximum rating of $4$ (``Extremely''). This implies that while participants distinguish screenshots with and without dark patterns with a significant difference, based on the five characteristics, their rating is overall rather low.

\subsubsection{Per Characteristic Rating}

\begin{table}[h]
\begin{tabular}{p{.09\linewidth}p{0.12\linewidth}p{0.12\linewidth}p{0.12\linewidth}p{0.12\linewidth}p{0.12\linewidth}p{0.12\linewidth}p{0.12\linewidth}}
\toprule
\multicolumn{6}{c}{\textbf{Comparison of Five Characteristics}} \\ \midrule
\multicolumn{6}{c}{Dark Pattern Screenshots} \\ \midrule
 & \multicolumn{1}{c}{Asym-} & \multicolumn{1}{c}{Covert} & \multicolumn{1}{c}{Restric-} & \multicolumn{1}{c}{Decep-} & \multicolumn{1}{c}{Hides} \\
 & \multicolumn{1}{c}{metry} & \multicolumn{1}{c}{} & \multicolumn{1}{c}{tive} & \multicolumn{1}{c}{tive} & \multicolumn{1}{c}{Info.} \\ \midrule
\rowcolor[HTML]{EFEFEF} 
\multicolumn{1}{l|}{mean}      &\centering{1.42}   &\centering{1.21}   &\centering{1.40}   &\centering{1.02}   &\centering{1.27}   \cr
\multicolumn{1}{l|}{median}    &\centering\textbf{1.00}   &\centering\textbf{1.00}   &\centering\textbf{1.00}   &\centering\textbf{1.00}   &\centering\textbf{1.00}   \cr
\rowcolor[HTML]{EFEFEF} 
\multicolumn{1}{l|}{SD}        &\centering{1.26}   &\centering{1.20}   &\centering{1.18}   &\centering{1.18}   &\centering{1.26}   \cr 
\toprule
\multicolumn{6}{c}{Non-Dark Pattern Screenshots} \\ \midrule
\rowcolor[HTML]{EFEFEF} 
\multicolumn{1}{l|}{mean}      &\centering{0.71}   &\centering{0.80}   &\centering{0.84}   &\centering{0.60}   &\centering{0.80}   \cr
\multicolumn{1}{l|}{median}    &\centering\textbf{0.00}   &\centering\textbf{0.00}   &\centering\textbf{0.00}   &\centering\textbf{0.00}   &\centering\textbf{0.00}   \cr
\rowcolor[HTML]{EFEFEF} 
\multicolumn{1}{l|}{SD}        &\centering{1.03}   &\centering{1.08}   &\centering{1.12}   &\centering{0.99}   &\centering{1.11}   \cr
\bottomrule
\end{tabular}
\caption{Overview of the mean, median, and standard deviation of participants' ratings of dark pattern and non-dark pattern screenshots according to Mathur et al.'s~\cite{mathur2019} five characteristics: \textit{assymetric}, \textit{covert}, \textit{restrictive}, \textit{deceptive}, and \textit{information hiding.}}
\label{tab:per_characteristic}
\end{table}

%
%
%

\begin{table}[h]
\begin{tabular}{p{.05\linewidth}p{0.05\linewidth}p{0.05\linewidth}p{0.05\linewidth}p{0.05\linewidth}p{0.07\linewidth}p{0.07\linewidth}p{0.07\linewidth}p{0.07\linewidth}}
\toprule
\multicolumn{9}{c}{\textbf{Comparison Of Screenshots}} \\ \midrule
\multicolumn{9}{c}{Dark Pattern Screenshots} \\ \midrule
\textbf{} & \multicolumn{1}{l}{\small{F1}} & \multicolumn{1}{l}{\small{F2}} & \multicolumn{1}{l}{\small{I1}} & \multicolumn{1}{l}{\small{I2}} & \multicolumn{1}{l}{\small{Ti1}} & \multicolumn{1}{l}{\small{Ti2}} & \multicolumn{1}{l}{\small{Tw1}} & \multicolumn{1}{l}{\small{Tw2}}\\ \midrule
\rowcolor[HTML]{EFEFEF} 
\multicolumn{1}{l|}{mean}      &1.40   &1.42   &1.45   &1.21   &1.76   &1.14    &\textbf{0.60}   &1.12 \\
\multicolumn{1}{l|}{median}    &1.40   &1.40   &1.40   &1.20   &1.80   &1.00    &\textbf{0.40}   &1.00 \\
\rowcolor[HTML]{EFEFEF} 
\multicolumn{1}{l|}{SD}        &1.08   &0.94   &1.08   &0.99   &1.06   &0.99    &\textbf{0.73}   &0.89 \\
\midrule
\multicolumn{9}{c}{Non-Dark Pattern Screenshots} \\ \midrule
\textbf{} & \multicolumn{1}{l}{\small{FA}} & \multicolumn{1}{l}{\small{FB}} & \multicolumn{1}{l}{\small{IA}} & \multicolumn{1}{l}{\small{IB}} & \multicolumn{1}{l}{\small{TiA}} & \multicolumn{1}{l}{\small{TiB}} & \multicolumn{1}{l}{\small{TwA}} & \multicolumn{1}{l}{\small{TwB}}\\ \midrule
\rowcolor[HTML]{EFEFEF} 
\multicolumn{1}{l|}{mean}      &\textbf{1.06}   &0.66   &0.45   &0.54   &0.69   &\textbf{1.10}    &0.39   &0.56 \\
\multicolumn{1}{l|}{median}    &\textbf{1.00}   &0.20   &0.00   &0.20   &0.40   &\textbf{1.00}    &0.00   &0.20 \\
\rowcolor[HTML]{EFEFEF} 
\multicolumn{1}{l|}{SD}        &\textbf{0.99}   &0.92   &0.71   &0.73   &0.81   &\textbf{0.99}    &0.65   &0.75 \\
\end{tabular}
\caption{Overview of the mean, median, and standard deviation of participants' ratings per dark pattern and non-dark pattern screenshot. Each of the four SNSs was represented with two screenshots containing dark patterns and two that did not. The letters in the screenshots' labels refer to a particular SNS: F = Facebook; I = Instagram; Ti = TikTok; Tw = Twitter. 
}
\label{tab:per_screenshot}
\end{table}

Based on participants' different ratings for dark pattern versus non-dark pattern screenshots, we gain a more detailed view of the applicability of the individual characteristics. We consider the median scores here because the data is not normally distributed. Overall, the median data indicates that across screenshots of the same kind, each characteristic contributed to the assessment, with a rating of 1 for screenshots that contain dark patterns and 0 for those not featuring dark patterns. 

To further validate the five characteristics, we investigated their relationship to the malice rating from section~\ref{sec:results-rec-dp}. We performed a multiple linear regression to see how well the individual characteristics predict the malice rating. The result shows a F-statistic p-value of < $0.0001$, suggesting that at least one of the five characteristics is significantly related to the malice score. Considering each t-statics, further analysis revealed that the characteristics \textit{asymmetric} (t < $0.001$) and \textit{restrictive} (t = $0.004$) show a significant association with the malice score. The remaining characteristics \textit{covert} (t = $0.053$), \textit{deceptive} (t = $0.081$), and \textit{hides information} (t = $0.074$) do not yield such association, however.
Thus, changes in those three characteristics do not significantly affect the malice score in our model.
\subsubsection{Per Screenshot Rating}

Considering the screenshots independently, we gain further insights into the differences between average scores. This allows us to notice the effectiveness and sensitivity with which this approach measures the malice in a single screenshot. Across the eight screenshots containing dark patterns, seven screenshots have median ratings >1, while the median rating for one screenshot is 0.4 (see Table~\ref{tab:per_screenshot}, Tw1). Looking at the non-dark pattern screenshots, six were rated with a median <1, while two screenshots have a median rating of 1 (see Table~\ref{tab:per_screenshot}, F1 and Ti2).

\section{Discussion}
This work presents insights from two studies, widening our understanding of how dark patterns manifest in SNSs and exploring a novel approach to evaluate the malice of interfaces. As online regulations have been shown to lack protection of users~\cite{bowyer_human-gdpr_2022}, we were interested in the effectiveness of current regulations that aim to shield users from dark patterns. Based on a comprehensive taxonomy, we let experienced HCI researchers apply dark patterns, by means of their descriptions, to four popular SNSs (Facebook, Instagram, TikTok, and Twitter). Although a range of dark patterns has been recognised, the results of the first study bear certain difficulties that hindered the process and thus highlight a necessity for more efficient approaches to recognising dark patterns. Exploring an alternative approach to evaluate the malice of interfaces, we defined five questions based on Mathur et al.'s~\cite{mathur2019} dark pattern characteristics. Letting regular users rate screenshots sampled from recordings of the first study, we found a potential measure in this approach that can be of aid for regulatory strategies. In this section, we discuss the applicability of dark pattern research as a tool to evaluate interfaces in relation to regulation.

\subsection{A Taxonomy As Evaluation Tool}
We acknowledge that the applied taxonomy, including entailed dark patterns from eight works, was not designed as a tool for the assessment of dark patterns and covers different scopes regarding their level of abstraction. While research on dark patterns moves forward, expanding our knowledge of the types of dark patterns that exist, we believe that it is important to reflect on the current status quo and consider the multitude of findings in new contexts. Study 1, therefore, tests the utility of dark patterns to identify their instances in SNSs. With the successful recognition of a range of these dark patterns in SNSs, the results of our first study imply that the chosen approach is suitable for identifying dark patterns in domains that may lie outside their original scope, offering an answer to our first research question. Tainting these results, however, we noticed certain issues that posed difficulties to the reviewers when executing their tasks. 

Overall, 31 out of 69 considered dark patterns were recognised, leaving another 31 not applicable in the context of SNSs. Especially game-related dark patterns~\cite{zagal_dark_2013} and those inspired by proxemic theory~\cite{greenberg_dark_nodate} were not all or rarely noticed. In contrast, dark patterns by Gray et al.~\cite{gray2018} were identified more frequently. This implies that expert reviewers found it easier to recognise dark patterns that were described more abstractly compared to domain-specific ones suggesting similar effectiveness in identifying dark patterns in regulatory contexts. 
A particular difficulty in this study emerged from dark patterns that shared the same names. Brignull's~\cite{brignull_deceptive_nodate} \textit{Confirmshaming} dark pattern, for instance, was carried over by Mathur et al.~\cite{mathur2019} who remained with its original definition, making it confusing as to which version should be applied when a related dark pattern is recognised. Other candidates - \textit{Privacy Zuckering} by Brignull~\cite{brignull_deceptive_nodate} and Bösch et al.~\cite{bosch2016} and \textit{Bait and Switch} by Brignull~\cite{brignull_deceptive_nodate} and Greenberg et al.~\cite{greenberg_dark_nodate} - were given distinct descriptions resulting in different applicability in SNSs. Contrary to this difficulty, the results of our co-occurrence tests show that dark patterns with different names apply in same interfaces. We see two possible explanations for this: (1) Provided descriptions of two dark patterns are too close, clouding distinct applications, at least in the context of SNSs. A high co-occurrence between \textit{Interface Interference}~\cite{gray2018} and \textit{Visual Interference}~\cite{mathur2019} can be explained this way. Alternatively, (2) two different dark patterns complement each other creating particularly problematic situations. Here, \textit{Privacy Zuckering} and \textit{Bad Default} do not describe the same interface problems but \textit{Privacy Zuckering} profits from the \textit{Bad Default} dark pattern as the latter will often result in users sharing more data unknowingly.

\subsection{Assessing the Malice of Interfaces}
\begin{figure*}
    \centering
    \includegraphics[width=.95\textwidth]{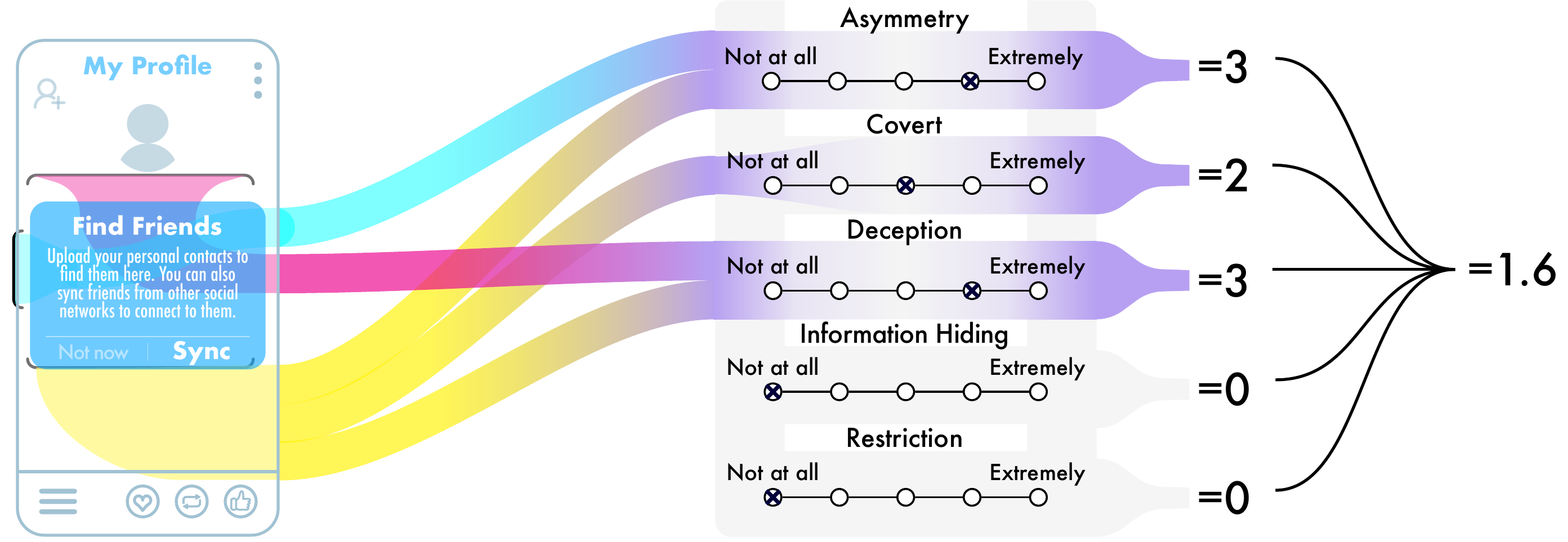}
    \caption{This figure demonstrates the approach to assess malice in interfaces by applying questions based on Mathur et al.'s~\cite{mathur2019} dark pattern characteristics. First, an interface is selected which is suspected of containing any amount of dark patterns. Using the five questions described in Table~\ref{tab:mathur2019}, the interface can then be evaluated using a Likert-scale from ``Not at all'' to ``Extremely''. In this example, we demonstrate this based on a five-item scale. The result are independent ratings for each characteristic, which can be averaged into a single digit.}
    \Description[Figure demonstrating our approach to assess the malice in interfaces.]{This figure demonstrates the approach to assess malice in interfaces by applying questions based on Mathur et al.'s (2019) dark pattern characteristics. On the left-hand side, a fictive illustration of a mobile screenshot is visible that contains abstractions of dark patterns. From each dark pattern, a coloured Sankey curve emerges, connecting the interface element to corresponding characteristics adopted from Mathur et al. (2019). One curve connects to the asymmetry characteristic, another one to the covert characteristics, and the third connects to deception. At the end of the Sankey curves are the five characteristics, each connected to a 5-point Likert-scale ranging from not at all to extremely. Fictional ratings for the characteristics are: asymmetry = 3, covert = 2, deception = 3, and information hiding and restriction are both rated 0. Five lines emerge behind each score, illustrating their combination into a single average of 1.6.}
    \label{fig:dp_vectorisation}
\end{figure*}

The results of study 1 indicate that abstract and distinct criteria are most efficient for evaluating the presence of dark patterns in interfaces. Study 2, therefore, explores an alternative approach by relying on Mathur et al.'s~\cite{mathur2019} five high-level characteristics to assess the malice of interfaces. Based on their framework, we developed five questions that we used to study regular users' ability to recognise dark patterns based on screenshots of the four SNSs. Answering our second research question, the results of this second study show that users were generally able to distinguish between screenshots featuring dark patterns and those that did not. However, ratings for the dark pattern screenshots indicate some difficulties as scores were considerably low (average median = 1.2), given that the maximum score a screenshot could receive is 4. Yet, participants' ability to differentiate screenshots based on these five characteristics suggests the promising effectiveness of this approach. Past work has found difficulties among participants in avoiding dark patterns~\cite{digeronimo2020, maier_dark_2020}. While our data suggest similar difficulties, our second study's results further support suggestions by Bongard-Blanchy et al.~\cite{BongardBlanchy2021}, who have shown that informing users about dark patterns helps to identify them.


This is further supported by the median ratings of each evaluated characteristic of the sixteen screenshots. We notice that across the eight dark pattern screenshots, each rating is 1 (``A little bit''), whereas the median rating for non-dark pattern screenshots is 0 (``Not at all''), as shown in Table~\ref{tab:per_characteristic}. This consistency across participants implies that all characteristics contribute to the assessment of dark patterns in screenshots. Considering individual median ratings per screenshot (see Table~\ref{tab:per_screenshot}), we see this consistency almost entirely confirmed. With regards to the dark pattern screenshots, participants were able to correctly identify malicious interfaces in seven out of eight instances (87.5\%). In non-dark pattern screenshots, participants accurately determined no presence of dark patterns six out of eight times (75\%). 
As neither the taxonomy nor Mathur et al.'s~\cite{mathur2019} characteristics were designed to identify or recognise dark patterns in SNSs, this attempt opens a possible pathway for future directions of dark pattern research. Relying on more abstract characteristics offers a promising approach to evaluating new interfaces. Figure~\ref{fig:dp_vectorisation} visually demonstrates this approach. If an interface is suspected of containing any number of dark patterns, it is evaluated using a 5-point Likert-scale (``Not at all'' - ``Extremely'') according to the five questions adopting Mathur et al.'s~\cite{mathur2019} characteristics. The maliciousness of the interface can then be determined by considering each characteristic's rating based on their individual values or as an average calculated from all five. We gain further support for this model through the multiple linear regression showing a highly significant relationship between the questions and the malice score. Individually, two characteristics -- \textit{asymmetry} and \textit{restrictive} -- maintain this highly significant association while three do not, leaving room for future improvement. The nature of this study describes an experimental setup aiming to assess the malice in interfaces better. The general statistical significance of both users' ability to differentiate between malicious and harmless design as well as in our multiple linear regression affirms the utility of such characteristics and our model.
This approach allows further insights into the types of dark patterns present in the interface by considering which characteristics they subscribe to. As participants of the second study only had to meet the criteria of being regular users of SNSs, we believe that more experienced evaluators could be able to evaluate interfaces more sensitively. 
Although this work utilises a total of 69 types of dark patterns, we acknowledge that our work has left new gaps for future work to consider SNS-specific types of dark patterns. Meanwhile, recent efforts have extended our knowledge of dark patterns in SNSs~\cite{gunawan_comparative_2021, schaffner_understanding_2022, habib_identifying_2022, mildner2022}, which leaves room for future updates of our research.
However, while these prior efforts describe dark patterns that occur in SNSs based on qualitative approaches, to our knowledge, this research is among the first to quantitatively assess dark patterns in SNSs while considering both experts' and users' ability to recognise them in this environment. Moreover, we extend the current discourse with a possible measure to access the malice of interfaces, regardless of their origin, by not requiring a complete corpus after all. Instead, relying on wider characteristics enables users to assess this malice based on five simple yet extendable, high-level dimensions.

\subsection{Paving The Way For Regulations} 
The variety of dark pattern types shows how far-stretched mischievous strategies in online domains can be. Still, they all have one thing in common: They harm users. Regulators and legislation already have powerful tools to ensure the protection of end-users. However, not all regulations are equally effective. To support this, findings from HCI research on dark patterns can aid existing approaches to protect peoples’ privacy on problematic designs.
The presented work has mainly two implications for legislative efforts regarding dark patterns.
The first one addresses the problem that the law is prone to lag behind dark patterns evolution, suggesting alternative approaches are needed to protect users successfully. The regulation of dark patterns must, on the one hand, be concrete enough to address manipulative mechanisms and, on the other hand, abstract enough to capture future developments. Our findings show that research in HCI constantly explores new dark patterns resulting in diverse taxonomies, as depicted in Figure~\ref{tab:dp_taxonomy}. Nevertheless, we see that recognising dark pattern characteristics on a meta-level is convincing and, referring to Mathur et al.'s high-level characteristics~\cite{mathur2019}, might be a promising approach to achieving a shared conceptualisation. This suggests that generalisable definitions and characterisations are better suited and more future-proof to assess dark patterns in various domains. We argue that findings from HCI can support legislative efforts by providing dark pattern characteristics based on empirical research and offering a sustainable vocabulary helping lawmakers to get ahead of developments of unethical designs. Such characteristics could be a basis for a legal definition and a general ban on dark patterns.
The second implication deals with recognising dark patterns in practice. Tools from HCI have the compelling potential for supporting courts and authorities since they could objectively measure the manipulation effect of a design (e.g. Figure~\ref{fig:dp_vectorisation}). Offering authorities a tool to evaluate the malice of interfaces easily, the proposed score determines the degree to which a specific design is either harmless or contains malicious features based on empirical evidence. Here, the goal lies in the identification of a certain score within the sweet spot, or threshold, that most accurately distinguishes between interfaces with dark patterns from those without. Our results show that even regular users are able to correctly differentiate between malicious and harmless interfaces. Professionals and trained people would likely perform similar tasks with even better accuracy. Consequently, the findings and tools from HCI research can become a considerable and valuable instrument in the decision-making processes of authorities. Ultimately, HCI research can pave the way for regulators to act on observed exploitation in interfaces that can, but are not limited to, target users’ personal data or manipulate their decision space, provoking potentially harmful actions.


\section{Limitations \& Future Work}
Both studies of this work yield certain limitations. Firstly, study 1 was conducted during the COVID-19 pandemic, which meant that the experiment was conducted without supervision. Although recordings do not suggest misunderstandings across reviewers, a present study supervisor can offer additional assistance. While we aimed to consider a range of SNSs, the number of platforms available today limited us to four applications with similar functionalities. Although the chosen SNSs present popular platforms, we neglected important services like YouTube or Twitch, featuring video-streaming platforms, but also messenger services like WhatsApp or Telegram, which each entail large user bases. Future work could consider alternative SNSs that were not in the scope of this work. As Mathur et al.'s~\cite{mathur2021} sixth \textit{Disparate Treatment} characteristic was not applied at all during the reviews, meaning that none of Zagal et al.'s~\cite{zagal_dark_2013} dark patterns were recognised in SNSs, it would further be interesting to consider SNSs that offer paying users different experiences (e.g. LinkedIn, Twitch, or YouTube). Also, future work could include recording instances of users sharing their data in- and outside of SNSs, as we did not include such a task in our cognitive walkthroughs. Study 1 was further limited by the selection of dark patterns included in our taxonomy. Because we decided only to include dark patterns that resulted from empirical research, we excluded those part of guidelines and regulations. Furthermore, Gunawan et al.~\cite{gunawan_comparative_2021} propose twelve additional dark patterns that we did not include as our experiment was conducted at the time of their publication. Future work could include further types of dark patterns for gaining an even deeper understanding of dark patterns in SNSs. Moreover, our methodology proved fruitful gaining us important insights into dark patterns in SNSs. Future work could adopt this approach to utilising the existing corpus of dark pattern knowledge when investigating dark patterns in other domains.

In study 2, we tested our evaluation approach based on screenshots to assess the malice of interfaces. While results indicate certain accuracy in participants differentiating between screenshots containing dark patterns and those that do not, our results do not allow us to make any statements about how well participants identified specific dark patterns. Furthermore, the screenshots are limited to showing dark patterns within a single stage on a static image. While we made sure to choose dark patterns, which are recognisable on screenshots, this limitation excludes possible dark patterns that rather work on a procedural level during an interaction. To reach participants, we used the online research platform \textit{Prolific}~\cite{prolific} to generate a convenience sample, restricted only to users who have prior experience with SNSs and are fluent in the English language, as screenshots were in English. However, we did not aim for a representative sample. Surprisingly, we noticed that 80,3\% of the participants identified as females skewing the demographic. Although we did not notice any differences between individual participants' ratings, we acknowledge that the data set is biased towards females. Moreover, we decided to rely on regular users as participants for this study. As our findings suggest a novel approach to aid the regulation of dark patterns, it would be interesting the see how related professionals such as regulators and legal scholars recognise dark patterns in a similar study. This could further be enhanced by additional characteristics that better incorporate malicious interfaces currently not covered. Also, Gunawan et al.~\cite{gunawan_comparative_2021} suggest that dark patterns may exist in SNSs to a different extent in their desktop modality. While we identified a host in SNSs for existing dark patterns, this work considers dark patterns that are not specific to this domain. As many described dark patterns have their origin in online shopping websites, future work could investigate social media platforms to describe unique dark patterns here. This further includes the characteristics from Mathur et al.~\cite{mathur2019}, which we used in our survey. Although the results of the multiple linear regression indicate a highly significant relationship between the questions and the malice score, only two out of five characteristics also yielded significant associations. This invites future research to advance our model and develop a suitable questionnaire for improved assessment.


\section{Conclusion}
In this paper, we examined four popular SNS platforms (Facebook, Instagram, TikTok, and Twitter) for dark patterns, advancing research in this context. Based on a cognitive walkthrough with six HCI experts, we learned which dark patterns occur in SNSs by considering a taxonomy based on prior findings in this field. Results of this study show that while this approach offers detailed insights, it lacks certain efficiency while posing difficulties to reviewers. Considering these results, we designed a novel approach to assess the malice of interfaces based on high-level characteristics. In a second study, we tested this alternative demonstrating a tool to recognise dark patterns in screenshots. Taking a legal perspective on current regulations for dark patterns, we discuss the findings of our second study, shining a light on how HCI research can aid the protection of SNS users. 

\begin{acks}
The research of this work was partially supported by the Klaus Tschira Stiftung gGmbH.
\end{acks}

\bibliographystyle{ACM-Reference-Format}
\bibliography{references.bib}


\begin{thebibliography}{45}


\ifx \showCODEN    \undefined \def \showCODEN     #1{\unskip}     \fi
\ifx \showDOI      \undefined \def \showDOI       #1{#1}\fi
\ifx \showISBNx    \undefined \def \showISBNx     #1{\unskip}     \fi
\ifx \showISBNxiii \undefined \def \showISBNxiii  #1{\unskip}     \fi
\ifx \showISSN     \undefined \def \showISSN      #1{\unskip}     \fi
\ifx \showLCCN     \undefined \def \showLCCN      #1{\unskip}     \fi
\ifx \shownote     \undefined \def \shownote      #1{#1}          \fi
\ifx \showarticletitle \undefined \def \showarticletitle #1{#1}   \fi
\ifx \showURL      \undefined \def \showURL       {\relax}        \fi
\providecommand\bibfield[2]{#2}
\providecommand\bibinfo[2]{#2}
\providecommand\natexlab[1]{#1}
\providecommand\showeprint[2][]{arXiv:#2}

\bibitem[\protect\citeauthoryear{2021}{2021}{2021}]%
        {reddit_samplesize}
\bibfield{author}{\bibinfo{person}{Reddit Inc~© 2021}.}
  \bibinfo{year}{2021}\natexlab{}.
\newblock \bibinfo{title}{r/SampleSize: {\textbar} Where your opinions actually
  matter!}
\newblock
\newblock
\urldef\tempurl%
\url{https://www.reddit.com/r/SampleSize/}
\showURL{%
\tempurl}
\newblock
\shownote{(visited on 2021-08-25).}


\bibitem[\protect\citeauthoryear{Ahn and Shin}{Ahn and Shin}{2013}]%
        {ahn2013social}
\bibfield{author}{\bibinfo{person}{Dohyun Ahn} {and} \bibinfo{person}{Dong-Hee
  Shin}.} \bibinfo{year}{2013}\natexlab{}.
\newblock \showarticletitle{Is the social use of media for seeking
  connectedness or for avoiding social isolation? Mechanisms underlying media
  use and subjective well-being}.
\newblock \bibinfo{journal}{\emph{Computers in Human Behavior}}
  \bibinfo{volume}{29}, \bibinfo{number}{6} (\bibinfo{year}{2013}),
  \bibinfo{pages}{2453--2462}.
\newblock


\bibitem[\protect\citeauthoryear{Beyens, Pouwels, van Driel, Keijsers, and
  Valkenburg}{Beyens et~al\mbox{.}}{2020}]%
        {beyens2020effect}
\bibfield{author}{\bibinfo{person}{Ine Beyens}, \bibinfo{person}{J~Loes
  Pouwels}, \bibinfo{person}{Irene~I van Driel}, \bibinfo{person}{Loes
  Keijsers}, {and} \bibinfo{person}{Patti~M Valkenburg}.}
  \bibinfo{year}{2020}\natexlab{}.
\newblock \showarticletitle{The effect of social media on well-being differs
  from adolescent to adolescent}.
\newblock \bibinfo{journal}{\emph{Scientific Reports}} \bibinfo{volume}{10},
  \bibinfo{number}{1} (\bibinfo{year}{2020}), \bibinfo{pages}{1--11}.
\newblock


\bibitem[\protect\citeauthoryear{Bongard-Blanchy, Rossi, Rivas, Doublet,
  Koenig, and Lenzini}{Bongard-Blanchy et~al\mbox{.}}{2021}]%
        {BongardBlanchy2021}
\bibfield{author}{\bibinfo{person}{Kerstin Bongard-Blanchy},
  \bibinfo{person}{Arianna Rossi}, \bibinfo{person}{Salvador Rivas},
  \bibinfo{person}{Sophie Doublet}, \bibinfo{person}{Vincent Koenig}, {and}
  \bibinfo{person}{Gabriele Lenzini}.} \bibinfo{year}{2021}\natexlab{}.
\newblock \showarticletitle{”I Am Definitely Manipulated, Even When I Am
  Aware of It. It’s Ridiculous!” - Dark Patterns from the End-User
  Perspective}. In \bibinfo{booktitle}{\emph{Designing Interactive Systems
  Conference 2021}} (Virtual Event, USA) \emph{(\bibinfo{series}{DIS '21})}.
  \bibinfo{publisher}{Association for Computing Machinery},
  \bibinfo{address}{New York, NY, USA}, \bibinfo{pages}{763–776}.
\newblock
\showISBNx{9781450384766}
\urldef\tempurl%
\url{https://doi.org/10.1145/3461778.3462086}
\showDOI{\tempurl}


\bibitem[\protect\citeauthoryear{B{\"o}sch, Erb, Kargl, Kopp, and
  Pfattheicher}{B{\"o}sch et~al\mbox{.}}{2016}]%
        {bosch2016}
\bibfield{author}{\bibinfo{person}{Christoph B{\"o}sch},
  \bibinfo{person}{Benjamin Erb}, \bibinfo{person}{Frank Kargl},
  \bibinfo{person}{Henning Kopp}, {and} \bibinfo{person}{Stefan Pfattheicher}.}
  \bibinfo{year}{2016}\natexlab{}.
\newblock \showarticletitle{Tales from the Dark Side: Privacy Dark Strategies
  and Privacy Dark Patterns.}
\newblock \bibinfo{journal}{\emph{Proc. Priv. Enhancing Technol.}}
  \bibinfo{volume}{2016}, \bibinfo{number}{4} (\bibinfo{year}{2016}),
  \bibinfo{pages}{237--254}.
\newblock


\bibitem[\protect\citeauthoryear{Bowyer, Holt, Go~Jefferies, Wilson, Kirk, and
  David~Smeddinck}{Bowyer et~al\mbox{.}}{2022}]%
        {bowyer_human-gdpr_2022}
\bibfield{author}{\bibinfo{person}{Alex Bowyer}, \bibinfo{person}{Jack Holt},
  \bibinfo{person}{Josephine Go~Jefferies}, \bibinfo{person}{Rob Wilson},
  \bibinfo{person}{David Kirk}, {and} \bibinfo{person}{Jan David~Smeddinck}.}
  \bibinfo{year}{2022}\natexlab{}.
\newblock \showarticletitle{Human-{GDPR} {Interaction}: {Practical}
  {Experiences} of {Accessing} {Personal} {Data}}. In
  \bibinfo{booktitle}{\emph{{CHI} {Conference} on {Human} {Factors} in
  {Computing} {Systems}}}. \bibinfo{publisher}{ACM}, \bibinfo{address}{New
  Orleans LA USA}, \bibinfo{pages}{1--19}.
\newblock
\showISBNx{978-1-4503-9157-3}
\urldef\tempurl%
\url{https://doi.org/10.1145/3491102.3501947}
\showDOI{\tempurl}


\bibitem[\protect\citeauthoryear{Brignull}{Brignull}{2010}]%
        {brignull_deceptive_nodate}
\bibfield{author}{\bibinfo{person}{Harry Brignull}.}
  \bibinfo{year}{2010}\natexlab{}.
\newblock \bibinfo{title}{Deceptive {Design} – formerly darkpatterns.org}.
\newblock
\newblock
\urldef\tempurl%
\url{https://www.deceptive.design/}
\showURL{%
\tempurl}
\newblock
\shownote{Visited on 2022-03-29.}


\bibitem[\protect\citeauthoryear{Commission}{Commission}{2022}]%
        {eu_commission_2022_da}
\bibfield{author}{\bibinfo{person}{European Commission}.}
  \bibinfo{year}{2022}\natexlab{}.
\newblock \bibinfo{title}{Proposal for a regulation of the European Parliament
  and of the Council on harmonized rules on fair access to and use of data
  (Data Act)}.
\newblock
\newblock
\urldef\tempurl%
\url{https://eur-lex.europa.eu/legal-content/EN/TXT/PDF/?uri=CELEX:52022PC0068}
\showURL{%
\tempurl}


\bibitem[\protect\citeauthoryear{Conti and Sobiesk}{Conti and Sobiesk}{2010}]%
        {conti_malicious_2010}
\bibfield{author}{\bibinfo{person}{Gregory Conti} {and} \bibinfo{person}{Edward
  Sobiesk}.} \bibinfo{year}{2010}\natexlab{}.
\newblock \showarticletitle{Malicious interface design: exploiting the user}.
  In \bibinfo{booktitle}{\emph{Proceedings of the 19th international conference
  on {World} wide web - {WWW} '10}}. \bibinfo{publisher}{ACM Press},
  \bibinfo{address}{Raleigh, North Carolina, USA}, \bibinfo{pages}{271}.
\newblock
\showISBNx{978-1-60558-799-8}
\urldef\tempurl%
\url{https://doi.org/10.1145/1772690.1772719}
\showDOI{\tempurl}


\bibitem[\protect\citeauthoryear{Di~Geronimo, Braz, Fregnan, Palomba, and
  Bacchelli}{Di~Geronimo et~al\mbox{.}}{2020}]%
        {digeronimo2020}
\bibfield{author}{\bibinfo{person}{Linda Di~Geronimo}, \bibinfo{person}{Larissa
  Braz}, \bibinfo{person}{Enrico Fregnan}, \bibinfo{person}{Fabio Palomba},
  {and} \bibinfo{person}{Alberto Bacchelli}.} \bibinfo{year}{2020}\natexlab{}.
\newblock \showarticletitle{UI Dark Patterns and Where to Find Them: A Study on
  Mobile Applications and User Perception}. In
  \bibinfo{booktitle}{\emph{Proceedings of the 2020 CHI Conference on Human
  Factors in Computing Systems}} (Honolulu, HI, USA)
  \emph{(\bibinfo{series}{CHI '20})}. \bibinfo{publisher}{Association for
  Computing Machinery}, \bibinfo{address}{New York, NY, USA},
  \bibinfo{pages}{1–14}.
\newblock
\showISBNx{9781450367080}
\urldef\tempurl%
\url{https://doi.org/10.1145/3313831.3376600}
\showDOI{\tempurl}


\bibitem[\protect\citeauthoryear{Ernala, Burke, Leavitt, and Ellison}{Ernala
  et~al\mbox{.}}{2020}]%
        {ernala_how_2020}
\bibfield{author}{\bibinfo{person}{Sindhu~Kiranmai Ernala},
  \bibinfo{person}{Moira Burke}, \bibinfo{person}{Alex Leavitt}, {and}
  \bibinfo{person}{Nicole~B. Ellison}.} \bibinfo{year}{2020}\natexlab{}.
\newblock \showarticletitle{How {Well} {Do} {People} {Report} {Time} {Spent} on
  {Facebook}? {An} {Evaluation} of {Established} {Survey} {Questions} with
  {Recommendations}}.
\newblock In \bibinfo{booktitle}{\emph{Proceedings of the 2020 {CHI}
  {Conference} on {Human} {Factors} in {Computing} {Systems}}}.
  \bibinfo{publisher}{Association for Computing Machinery},
  \bibinfo{address}{New York, NY, USA}, \bibinfo{pages}{1--14}.
\newblock
\showISBNx{978-1-4503-6708-0}
\urldef\tempurl%
\url{https://doi.org/10.1145/3313831.3376435}
\showURL{%
\tempurl}


\bibitem[\protect\citeauthoryear{{European Parliament and the Council for the
  European Union}}{{European Parliament and the Council for the European
  Union}}{2016}]%
        {eu_gdpr_2016}
\bibfield{author}{\bibinfo{person}{{European Parliament and the Council for the
  European Union}}.} \bibinfo{year}{2016}\natexlab{}.
\newblock \bibinfo{title}{Regulation (EU) 2016/679 of the European Parliament
  and of the Council of 27 April 2016 on the protection of natural persons with
  regard to the processing of personal data and on the free movement of such
  data, and repealing Directive 95/46/EC (General Data Protection Regulation)
  (Text with EEA relevance)}.
\newblock
\newblock
\urldef\tempurl%
\url{https://eur-lex.europa.eu/legal-content/EN/TXT/?uri=CELEX\%3A32016R0679&qid=1684585264339}
\showURL{%
\tempurl}


\bibitem[\protect\citeauthoryear{{European Parliament and the Council for the
  European Union}}{{European Parliament and the Council for the European
  Union}}{2022}]%
        {eu_commission_2022_dsa}
\bibfield{author}{\bibinfo{person}{{European Parliament and the Council for the
  European Union}}.} \bibinfo{year}{2022}\natexlab{}.
\newblock \bibinfo{title}{Regulation (EU) 2022/2065 of the European Parliament
  and of the Council of 19 October 2022 on a Single Market For Digital Services
  and amending Directive 2000/31/EC (Digital Services Act) (Text with EEA
  relevance)}.
\newblock
\newblock
\urldef\tempurl%
\url{https://eur-lex.europa.eu/legal-content/EN/TXT/PDF/?uri=CELEX:32022R2065}
\showURL{%
\tempurl}


\bibitem[\protect\citeauthoryear{Faul, Erdfelder, Lang, and Buchner}{Faul
  et~al\mbox{.}}{2007}]%
        {gpower_2007}
\bibfield{author}{\bibinfo{person}{Franz Faul}, \bibinfo{person}{Edgar
  Erdfelder}, \bibinfo{person}{Albert-Georg Lang}, {and} \bibinfo{person}{Axel
  Buchner}.} \bibinfo{year}{2007}\natexlab{}.
\newblock \showarticletitle{G* Power 3: A flexible statistical power analysis
  program for the social, behavioral, and biomedical sciences}.
\newblock \bibinfo{journal}{\emph{Behavior research methods}}
  \bibinfo{volume}{39}, \bibinfo{number}{2} (\bibinfo{year}{2007}),
  \bibinfo{pages}{175--191}.
\newblock


\bibitem[\protect\citeauthoryear{Friese}{Friese}{2019}]%
        {friese_2019}
\bibfield{author}{\bibinfo{person}{Susanne Friese}.}
  \bibinfo{year}{2019}\natexlab{}.
\newblock \bibinfo{booktitle}{\emph{Qualitative data analysis with ATLAS. ti}
  (\bibinfo{edition}{3} ed.)}.
\newblock \bibinfo{publisher}{SAGE Publications Ltd},
  \bibinfo{address}{California, United States}. 344 pages.
\newblock
\showISBNx{9781526446237}


\bibitem[\protect\citeauthoryear{GmbH}{GmbH}{2021}]%
        {atlasti_2021}
\bibfield{author}{\bibinfo{person}{ATLAS.ti Scientific Software~Development
  GmbH}.} \bibinfo{year}{2021}\natexlab{}.
\newblock \bibinfo{title}{{ATLAS}.ti: {The} {Qualitative} {Data} {Analysis} \&
  {Research} {Software}}.
\newblock
\newblock
\urldef\tempurl%
\url{https://atlasti.com/}
\showURL{%
\tempurl}


\bibitem[\protect\citeauthoryear{Gray and Chivukula}{Gray and
  Chivukula}{2019}]%
        {gray_ethical_2019}
\bibfield{author}{\bibinfo{person}{Colin~M. Gray} {and}
  \bibinfo{person}{Shruthi~Sai Chivukula}.} \bibinfo{year}{2019}\natexlab{}.
\newblock \showarticletitle{Ethical {Mediation} in {UX} {Practice}}.
\newblock In \bibinfo{booktitle}{\emph{Proceedings of the 2019 {CHI}
  {Conference} on {Human} {Factors} in {Computing} {Systems}}}.
  \bibinfo{publisher}{Association for Computing Machinery},
  \bibinfo{address}{New York, NY, USA}, \bibinfo{pages}{1--11}.
\newblock
\showISBNx{978-1-4503-5970-2}
\urldef\tempurl%
\url{https://doi.org/10.1145/3290605.3300408}
\showURL{%
\tempurl}


\bibitem[\protect\citeauthoryear{Gray, Chivukula, and Lee}{Gray
  et~al\mbox{.}}{2020}]%
        {Gray2020a}
\bibfield{author}{\bibinfo{person}{Colin~M. Gray}, \bibinfo{person}{Shruthi~Sai
  Chivukula}, {and} \bibinfo{person}{Ahreum Lee}.}
  \bibinfo{year}{2020}\natexlab{}.
\newblock \bibinfo{booktitle}{\emph{What Kind of Work Do "Asshole Designers"
  Create? Describing Properties of Ethical Concern on Reddit}}.
\newblock \bibinfo{publisher}{Association for Computing Machinery},
  \bibinfo{address}{New York, NY, USA}, \bibinfo{pages}{61–73}.
\newblock
\showISBNx{9781450369749}
\urldef\tempurl%
\url{https://doi.org/10.1145/3357236.3395486}
\showURL{%
\tempurl}


\bibitem[\protect\citeauthoryear{Gray, Kou, Battles, Hoggatt, and Toombs}{Gray
  et~al\mbox{.}}{2018}]%
        {gray2018}
\bibfield{author}{\bibinfo{person}{Colin~M. Gray}, \bibinfo{person}{Yubo Kou},
  \bibinfo{person}{Bryan Battles}, \bibinfo{person}{Joseph Hoggatt}, {and}
  \bibinfo{person}{Austin~L. Toombs}.} \bibinfo{year}{2018}\natexlab{}.
\newblock \bibinfo{booktitle}{\emph{The Dark (Patterns) Side of UX Design}}.
\newblock \bibinfo{publisher}{Association for Computing Machinery},
  \bibinfo{address}{New York, NY, USA}, \bibinfo{pages}{1–14}.
\newblock
\showISBNx{9781450356206}
\urldef\tempurl%
\url{https://doi.org/10.1145/3173574.3174108}
\showURL{%
\tempurl}


\bibitem[\protect\citeauthoryear{Gray, Santos, Bielova, Toth, and
  Clifford}{Gray et~al\mbox{.}}{2021}]%
        {Gray2021}
\bibfield{author}{\bibinfo{person}{Colin~M. Gray}, \bibinfo{person}{Cristiana
  Santos}, \bibinfo{person}{Nataliia Bielova}, \bibinfo{person}{Michael Toth},
  {and} \bibinfo{person}{Damian Clifford}.} \bibinfo{year}{2021}\natexlab{}.
\newblock \bibinfo{booktitle}{\emph{Dark Patterns and the Legal Requirements of
  Consent Banners: An Interaction Criticism Perspective}}.
\newblock \bibinfo{publisher}{Association for Computing Machinery},
  \bibinfo{address}{New York, NY, USA}, \bibinfo{pages}{pp. 1--18}.
\newblock
\showISBNx{9781450380966}
\urldef\tempurl%
\url{https://doi.org/10.1145/3411764.3445779}
\showURL{%
\tempurl}


\bibitem[\protect\citeauthoryear{Graßl, Schraffenberger, Borgesius, and
  Buijzen}{Graßl et~al\mbox{.}}{2021}]%
        {grasl_dark_2021}
\bibfield{author}{\bibinfo{person}{Paul Graßl}, \bibinfo{person}{Hanna
  Schraffenberger}, \bibinfo{person}{Frederik~Zuiderveen Borgesius}, {and}
  \bibinfo{person}{Moniek Buijzen}.} \bibinfo{year}{2021}\natexlab{}.
\newblock \showarticletitle{Dark and {Bright} {Patterns} in {Cookie} {Consent}
  {Requests}}.
\newblock \bibinfo{journal}{\emph{Journal of Digital Social Research}}
  \bibinfo{volume}{3}, \bibinfo{number}{1} (\bibinfo{date}{Feb.}
  \bibinfo{year}{2021}), \bibinfo{pages}{1--38}.
\newblock
\showISSN{2003-1998}
\urldef\tempurl%
\url{https://doi.org/10.33621/jdsr.v3i1.54}
\showDOI{\tempurl}
\newblock
\shownote{Number: 1.}


\bibitem[\protect\citeauthoryear{Greenberg, Boring, Vermeulen, and
  Dostal}{Greenberg et~al\mbox{.}}{2014}]%
        {greenberg_dark_nodate}
\bibfield{author}{\bibinfo{person}{Saul Greenberg}, \bibinfo{person}{Sebastian
  Boring}, \bibinfo{person}{Jo Vermeulen}, {and} \bibinfo{person}{Jakub
  Dostal}.} \bibinfo{year}{2014}\natexlab{}.
\newblock \bibinfo{booktitle}{\emph{Dark Patterns in Proxemic Interactions: A
  Critical Perspective}}.
\newblock \bibinfo{publisher}{Association for Computing Machinery},
  \bibinfo{address}{New York, NY, USA}, \bibinfo{pages}{523–532}.
\newblock
\showISBNx{9781450329026}
\urldef\tempurl%
\url{https://doi.org/10.1145/2598510.2598541}
\showDOI{\tempurl}


\bibitem[\protect\citeauthoryear{Gunawan, Pradeep, Choffnes, Hartzog, and
  Wilson}{Gunawan et~al\mbox{.}}{2022}]%
        {gunawan_comparative_2021}
\bibfield{author}{\bibinfo{person}{Johanna Gunawan}, \bibinfo{person}{Amogh
  Pradeep}, \bibinfo{person}{David Choffnes}, \bibinfo{person}{Woodrow
  Hartzog}, {and} \bibinfo{person}{Christo Wilson}.}
  \bibinfo{year}{2022}\natexlab{}.
\newblock \showarticletitle{A Comparative Study of Dark Patterns Across Web and
  Mobile Modalities}.
\newblock   \bibinfo{volume}{5} (\bibinfo{year}{2022}), \bibinfo{pages}{1--29}.
\newblock
Issue {CSCW}2.
\showISSN{2573-0142}
\urldef\tempurl%
\url{https://doi.org/10.1145/3479521}
\showDOI{\tempurl}


\bibitem[\protect\citeauthoryear{Habib, Pearman, Young, Saxena, Zhang, and
  Cranor}{Habib et~al\mbox{.}}{2022}]%
        {habib_identifying_2022}
\bibfield{author}{\bibinfo{person}{Hana Habib}, \bibinfo{person}{Sarah
  Pearman}, \bibinfo{person}{Ellie Young}, \bibinfo{person}{Ishika Saxena},
  \bibinfo{person}{Robert Zhang}, {and} \bibinfo{person}{Lorrie~{FaIth}
  Cranor}.} \bibinfo{year}{2022}\natexlab{}.
\newblock \showarticletitle{Identifying User Needs for Advertising Controls on
  Facebook}.
\newblock   \bibinfo{volume}{6} (\bibinfo{year}{2022}), \bibinfo{pages}{1--42}.
\newblock
Issue {CSCW}1.
\showISSN{2573-0142}
\urldef\tempurl%
\url{https://doi.org/10.1145/3512906}
\showDOI{\tempurl}


\bibitem[\protect\citeauthoryear{Hustinx}{Hustinx}{2010}]%
        {hustinx_privacy_2010}
\bibfield{author}{\bibinfo{person}{Peter Hustinx}.}
  \bibinfo{year}{2010}\natexlab{}.
\newblock \showarticletitle{Privacy by design: delivering the promises}.
\newblock \bibinfo{journal}{\emph{Identity in the Information Society}}
  \bibinfo{volume}{3}, \bibinfo{number}{2} (\bibinfo{date}{Aug.}
  \bibinfo{year}{2010}), \bibinfo{pages}{253--255}.
\newblock
\showISSN{1876-0678}
\urldef\tempurl%
\url{https://doi.org/10.1007/s12394-010-0061-z}
\showDOI{\tempurl}


\bibitem[\protect\citeauthoryear{Jaspers, Steen, van~den Bos, and
  Geenen}{Jaspers et~al\mbox{.}}{2004}]%
        {jaspers2004}
\bibfield{author}{\bibinfo{person}{Monique~W.M. Jaspers},
  \bibinfo{person}{Thiemo Steen}, \bibinfo{person}{Cor van~den Bos}, {and}
  \bibinfo{person}{Maud Geenen}.} \bibinfo{year}{2004}\natexlab{}.
\newblock \showarticletitle{The think aloud method: a guide to user interface
  design}.
\newblock \bibinfo{journal}{\emph{International Journal of Medical
  Informatics}} \bibinfo{volume}{73}, \bibinfo{number}{11}
  (\bibinfo{year}{2004}), \bibinfo{pages}{781--795}.
\newblock
\showISSN{1386-5056}
\urldef\tempurl%
\url{https://doi.org/10.1016/j.ijmedinf.2004.08.003}
\showDOI{\tempurl}


\bibitem[\protect\citeauthoryear{Junco}{Junco}{2013}]%
        {junco_comparing_2013}
\bibfield{author}{\bibinfo{person}{Reynol Junco}.}
  \bibinfo{year}{2013}\natexlab{}.
\newblock \showarticletitle{Comparing actual and self-reported measures of
  {Facebook} use}.
\newblock \bibinfo{journal}{\emph{Computers in Human Behavior}}
  \bibinfo{volume}{29}, \bibinfo{number}{3} (\bibinfo{date}{May}
  \bibinfo{year}{2013}), \bibinfo{pages}{626--631}.
\newblock
\showISSN{0747-5632}
\urldef\tempurl%
\url{https://doi.org/10.1016/j.chb.2012.11.007}
\showDOI{\tempurl}


\bibitem[\protect\citeauthoryear{Kommission, und Verbraucher,
  Lupiáñez-Villanueva, Boluda, Bogliacino, Liva, Lechardoy, and Rodríguez
  de~las Heras~Ballell}{Kommission et~al\mbox{.}}{2022}]%
        {eu-2022-behaviour}
\bibfield{author}{\bibinfo{person}{Europäische Kommission},
  \bibinfo{person}{Generaldirektion~Justiz und Verbraucher}, \bibinfo{person}{F
  Lupiáñez-Villanueva}, \bibinfo{person}{A Boluda}, \bibinfo{person}{F
  Bogliacino}, \bibinfo{person}{G Liva}, \bibinfo{person}{L Lechardoy}, {and}
  \bibinfo{person}{T Rodríguez de~las Heras~Ballell}.}
  \bibinfo{year}{2022}\natexlab{}.
\newblock \bibinfo{booktitle}{\emph{Behavioural study on unfair commercial
  practices in the digital environment : dark patterns and manipulative
  personalisation : final report}}.
\newblock \bibinfo{publisher}{Amt für Veröffentlichungen der Europäischen
  Union}.
\newblock
\urldef\tempurl%
\url{https://doi.org/10.2838/859030}
\showDOI{\tempurl}


\bibitem[\protect\citeauthoryear{Legislature}{Legislature}{2018}]%
        {ccpa_2018}
\bibfield{author}{\bibinfo{person}{California~State Legislature}.}
  \bibinfo{year}{2018}\natexlab{}.
\newblock \bibinfo{title}{CCPA-18 2018. California Consumer Privacy Act of 2018
  [1798.100 - 1798.199] (CCPA).}
\newblock
\newblock
\urldef\tempurl%
\url{https://leginfo.legislature.ca.gov/faces/codes_displayText.xhtml?division=3.&part=4.&lawCode=CIV&title=1.81.5.}
\showURL{%
\tempurl}


\bibitem[\protect\citeauthoryear{Leiser and Caruana}{Leiser and
  Caruana}{2021}]%
        {leiser_dark_2021}
\bibfield{author}{\bibinfo{person}{M Leiser} {and} \bibinfo{person}{M
  Caruana}.} \bibinfo{year}{2021}\natexlab{}.
\newblock \showarticletitle{Dark {Patterns}: {Light} to be found in
  {Europe}’s {Consumer} {Protection} {Regime}}.
\newblock \bibinfo{journal}{\emph{Journal Of European Consumer And Market Law}}
   \bibinfo{volume}{10(6)} (\bibinfo{year}{2021}), \bibinfo{pages}{237--251}.
\newblock
\newblock
\shownote{Retrieved from https://hdl.handle.net/1887/3278362.}


\bibitem[\protect\citeauthoryear{Ltd}{Ltd}{2021}]%
        {prolific}
\bibfield{author}{\bibinfo{person}{Prolific~Academic Ltd}.}
  \bibinfo{year}{2021}\natexlab{}.
\newblock \bibinfo{title}{Prolific {\textbar} {Online} participant recruitment
  for surveys and market research}.
\newblock
\newblock
\urldef\tempurl%
\url{https://prolific.co/}
\showURL{%
\tempurl}
\newblock
\shownote{(visited on 2021-08-25).}


\bibitem[\protect\citeauthoryear{Maier}{Maier}{2020}]%
        {maier_dark_2020}
\bibfield{author}{\bibinfo{person}{Maximilian Maier}.}
  \bibinfo{year}{2020}\natexlab{}.
\newblock \showarticletitle{Dark {Design} {Patterns} - {An} {End}-user
  {Perspective}}.
\newblock \bibinfo{journal}{\emph{Human Technology}}  \bibinfo{volume}{16}
  (\bibinfo{year}{2020}), \bibinfo{pages}{170--199}.
\newblock
\urldef\tempurl%
\url{https://doi.org/10.17011/ht/urn.202008245641}
\showDOI{\tempurl}


\bibitem[\protect\citeauthoryear{Mathur, Acar, Friedman, Lucherini, Mayer,
  Chetty, and Narayanan}{Mathur et~al\mbox{.}}{2019}]%
        {mathur2019}
\bibfield{author}{\bibinfo{person}{Arunesh Mathur}, \bibinfo{person}{Gunes
  Acar}, \bibinfo{person}{Michael~J. Friedman}, \bibinfo{person}{Elena
  Lucherini}, \bibinfo{person}{Jonathan Mayer}, \bibinfo{person}{Marshini
  Chetty}, {and} \bibinfo{person}{Arvind Narayanan}.}
  \bibinfo{year}{2019}\natexlab{}.
\newblock \showarticletitle{{Dark Patterns at Scale}}.
\newblock \bibinfo{journal}{\emph{Proceedings of the ACM on Human-Computer
  Interaction}} \bibinfo{volume}{3}, \bibinfo{number}{CSCW}
  (\bibinfo{date}{nov} \bibinfo{year}{2019}), \bibinfo{pages}{1--32}.
\newblock
\showISSN{2573-0142}
\urldef\tempurl%
\url{https://doi.org/10.1145/3359183}
\showDOI{\tempurl}


\bibitem[\protect\citeauthoryear{Mathur, Kshirsagar, and Mayer}{Mathur
  et~al\mbox{.}}{2021}]%
        {mathur2021}
\bibfield{author}{\bibinfo{person}{Arunesh Mathur}, \bibinfo{person}{Mihir
  Kshirsagar}, {and} \bibinfo{person}{Jonathan Mayer}.}
  \bibinfo{year}{2021}\natexlab{}.
\newblock \bibinfo{booktitle}{\emph{What Makes a Dark Pattern... Dark? Design
  Attributes, Normative Considerations, and Measurement Methods}}.
\newblock \bibinfo{publisher}{Association for Computing Machinery},
  \bibinfo{address}{New York, NY, USA}, \bibinfo{pages}{pp. 1--18}.
\newblock
\showISBNx{9781450380966}
\urldef\tempurl%
\url{https://doi.org/10.1145/3411764.3445610}
\showURL{%
\tempurl}


\bibitem[\protect\citeauthoryear{Mildner and Savino}{Mildner and
  Savino}{2021}]%
        {mildner_ethical_2021}
\bibfield{author}{\bibinfo{person}{Thomas Mildner} {and}
  \bibinfo{person}{Gian-Luca Savino}.} \bibinfo{year}{2021}\natexlab{}.
\newblock \showarticletitle{Ethical {User} {Interfaces}: {Exploring} the
  {Effects} of {Dark} {Patterns} on {Facebook}}.
\newblock In \bibinfo{booktitle}{\emph{Extended {Abstracts} of the 2021 {CHI}
  {Conference} on {Human} {Factors} in {Computing} {Systems}}}.
  \bibinfo{publisher}{Association for Computing Machinery},
  \bibinfo{address}{New York, NY, USA}, \bibinfo{pages}{1--7}.
\newblock
\showISBNx{978-1-4503-8095-9}
\urldef\tempurl%
\url{https://doi.org/10.1145/3411763.3451659}
\showURL{%
\tempurl}


\bibitem[\protect\citeauthoryear{Mildner, Savino, Doyle, Cowan, and
  Malaka}{Mildner et~al\mbox{.}}{2023}]%
        {mildner2022}
\bibfield{author}{\bibinfo{person}{Thomas Mildner}, \bibinfo{person}{Gian-Luca
  Savino}, \bibinfo{person}{Philip~R. Doyle}, \bibinfo{person}{Benjamin~R.
  Cowan}, {and} \bibinfo{person}{Rainer Malaka}.}
  \bibinfo{year}{2023}\natexlab{}.
\newblock \showarticletitle{About Engaging and Governing Strategies: A Thematic
  Analysis of Dark Patterns in Social Networking Services}. In
  \bibinfo{booktitle}{\emph{Proceedings of the 2023 CHI Conference on Human
  Factors in Computing Systems}} (Hamburg, Germany) \emph{(\bibinfo{series}{CHI
  '23})}. \bibinfo{publisher}{Association for Computing Machinery},
  \bibinfo{address}{New York, NY, USA}, Article \bibinfo{articleno}{192},
  \bibinfo{numpages}{15}~pages.
\newblock
\showISBNx{9781450394215}
\urldef\tempurl%
\url{https://doi.org/10.1145/3544548.3580695}
\showDOI{\tempurl}


\bibitem[\protect\citeauthoryear{Nouwens, Liccardi, Veale, Karger, and
  Kagal}{Nouwens et~al\mbox{.}}{2020}]%
        {nouwens_dark_2020}
\bibfield{author}{\bibinfo{person}{Midas Nouwens}, \bibinfo{person}{Ilaria
  Liccardi}, \bibinfo{person}{Michael Veale}, \bibinfo{person}{David Karger},
  {and} \bibinfo{person}{Lalana Kagal}.} \bibinfo{year}{2020}\natexlab{}.
\newblock \showarticletitle{Dark {Patterns} after the {GDPR}: {Scraping}
  {Consent} {Pop}-ups and {Demonstrating} their {Influence}}. In
  \bibinfo{booktitle}{\emph{Proceedings of the 2020 {CHI} {Conference} on
  {Human} {Factors} in {Computing} {Systems}}}. \bibinfo{publisher}{ACM},
  \bibinfo{address}{Honolulu HI USA}, \bibinfo{pages}{1--13}.
\newblock
\showISBNx{978-1-4503-6708-0}
\urldef\tempurl%
\url{https://doi.org/10.1145/3313831.3376321}
\showDOI{\tempurl}


\bibitem[\protect\citeauthoryear{Schaffner, Lingareddy, and Chetty}{Schaffner
  et~al\mbox{.}}{2022}]%
        {schaffner_understanding_2022}
\bibfield{author}{\bibinfo{person}{Brennan Schaffner}, \bibinfo{person}{Neha~A.
  Lingareddy}, {and} \bibinfo{person}{Marshini Chetty}.}
  \bibinfo{year}{2022}\natexlab{}.
\newblock \showarticletitle{Understanding {Account} {Deletion} and {Relevant}
  {Dark} {Patterns} on {Social} {Media}}.
\newblock \bibinfo{journal}{\emph{Proceedings of the ACM on Human-Computer
  Interaction}} \bibinfo{volume}{6}, \bibinfo{number}{CSCW2}
  (\bibinfo{date}{Nov.} \bibinfo{year}{2022}), \bibinfo{pages}{1--43}.
\newblock
\showISSN{2573-0142}
\urldef\tempurl%
\url{https://doi.org/10.1145/3555142}
\showDOI{\tempurl}


\bibitem[\protect\citeauthoryear{Schoenebeck}{Schoenebeck}{2014}]%
        {schoenebeck_giving_2014}
\bibfield{author}{\bibinfo{person}{Sarita~Yardi Schoenebeck}.}
  \bibinfo{year}{2014}\natexlab{}.
\newblock \showarticletitle{Giving up {Twitter} for {Lent}: how and why we take
  breaks from social media}. In \bibinfo{booktitle}{\emph{Proceedings of the
  {SIGCHI} {Conference} on {Human} {Factors} in {Computing} {Systems}}}
  \emph{(\bibinfo{series}{{CHI} '14})}. \bibinfo{publisher}{Association for
  Computing Machinery}, \bibinfo{address}{New York, NY, USA},
  \bibinfo{pages}{773--782}.
\newblock
\showISBNx{978-1-4503-2473-1}
\urldef\tempurl%
\url{https://doi.org/10.1145/2556288.2556983}
\showDOI{\tempurl}


\bibitem[\protect\citeauthoryear{Shakya and Christakis}{Shakya and
  Christakis}{2017}]%
        {shakya_association_2017}
\bibfield{author}{\bibinfo{person}{Holly~B. Shakya} {and}
  \bibinfo{person}{Nicholas~A. Christakis}.} \bibinfo{year}{2017}\natexlab{}.
\newblock \showarticletitle{Association of {Facebook} {Use} {With}
  {Compromised} {Well}-{Being}: {A} {Longitudinal} {Study}}.
\newblock \bibinfo{journal}{\emph{American Journal of Epidemiology}}
  \bibinfo{volume}{185}, \bibinfo{number}{3} (\bibinfo{date}{Feb.}
  \bibinfo{year}{2017}), \bibinfo{pages}{203--211}.
\newblock
\showISSN{0002-9262}
\urldef\tempurl%
\url{https://doi.org/10.1093/aje/kww189}
\showDOI{\tempurl}


\bibitem[\protect\citeauthoryear{Statista}{Statista}{2021}]%
        {statista_SNS_pop_2021}
\bibfield{author}{\bibinfo{person}{Statista}.} \bibinfo{year}{2021}\natexlab{}.
\newblock \bibinfo{title}{We Are Social, Hootsuite, DataReportal. (July 21,
  2021). Most popular social networks worldwide as of July 2021, ranked by
  number of active users (in millions)}.
\newblock
\newblock
\urldef\tempurl%
\url{https://www.statista.com/statistics/272014/global-social-networks-ranked-by-number-of-users/}
\showURL{%
\tempurl}


\bibitem[\protect\citeauthoryear{Utz, Degeling, Fahl, Schaub, and Holz}{Utz
  et~al\mbox{.}}{2019}]%
        {utz_informed_2019}
\bibfield{author}{\bibinfo{person}{Christine Utz}, \bibinfo{person}{Martin
  Degeling}, \bibinfo{person}{Sascha Fahl}, \bibinfo{person}{Florian Schaub},
  {and} \bibinfo{person}{Thorsten Holz}.} \bibinfo{year}{2019}\natexlab{}.
\newblock \showarticletitle{({Un})informed {Consent}: {Studying} {GDPR}
  {Consent} {Notices} in the {Field}}. In \bibinfo{booktitle}{\emph{Proceedings
  of the 2019 {ACM} {SIGSAC} {Conference} on {Computer} and {Communications}
  {Security}}} \emph{(\bibinfo{series}{{CCS} '19})}.
  \bibinfo{publisher}{Association for Computing Machinery},
  \bibinfo{address}{New York, NY, USA}, \bibinfo{pages}{973--990}.
\newblock
\showISBNx{978-1-4503-6747-9}
\urldef\tempurl%
\url{https://doi.org/10.1145/3319535.3354212}
\showDOI{\tempurl}


\bibitem[\protect\citeauthoryear{Wang, Jackson, Gaskin, and Wang}{Wang
  et~al\mbox{.}}{2014}]%
        {wang_effects_2014}
\bibfield{author}{\bibinfo{person}{Jin-Liang Wang}, \bibinfo{person}{Linda~A.
  Jackson}, \bibinfo{person}{James Gaskin}, {and} \bibinfo{person}{Hai-Zhen
  Wang}.} \bibinfo{year}{2014}\natexlab{}.
\newblock \showarticletitle{The effects of {Social} {Networking} {Site} ({SNS})
  use on college students’ friendship and well-being}.
\newblock \bibinfo{journal}{\emph{Computers in Human Behavior}}
  \bibinfo{volume}{37} (\bibinfo{date}{Aug.} \bibinfo{year}{2014}),
  \bibinfo{pages}{229--236}.
\newblock
\showISSN{0747-5632}
\urldef\tempurl%
\url{https://doi.org/10.1016/j.chb.2014.04.051}
\showDOI{\tempurl}


\bibitem[\protect\citeauthoryear{Wang, Norcie, Komanduri, Acquisti, Leon, and
  Cranor}{Wang et~al\mbox{.}}{2011}]%
        {wang2011}
\bibfield{author}{\bibinfo{person}{Yang Wang}, \bibinfo{person}{Gregory
  Norcie}, \bibinfo{person}{Saranga Komanduri}, \bibinfo{person}{Alessandro
  Acquisti}, \bibinfo{person}{Pedro~Giovanni Leon}, {and}
  \bibinfo{person}{Lorrie~Faith Cranor}.} \bibinfo{year}{2011}\natexlab{}.
\newblock \showarticletitle{"I Regretted the Minute I Pressed Share": A
  Qualitative Study of Regrets on Facebook}. In
  \bibinfo{booktitle}{\emph{Proceedings of the Seventh Symposium on Usable
  Privacy and Security}} (Pittsburgh, Pennsylvania)
  \emph{(\bibinfo{series}{SOUPS '11})}. \bibinfo{publisher}{Association for
  Computing Machinery}, \bibinfo{address}{New York, NY, USA}, Article
  \bibinfo{articleno}{10}, \bibinfo{numpages}{16}~pages.
\newblock
\showISBNx{9781450309110}
\urldef\tempurl%
\url{https://doi.org/10.1145/2078827.2078841}
\showDOI{\tempurl}


\bibitem[\protect\citeauthoryear{Zagal, Björk, and Lewis}{Zagal
  et~al\mbox{.}}{2013}]%
        {zagal_dark_2013}
\bibfield{author}{\bibinfo{person}{José~P Zagal}, \bibinfo{person}{Staffan
  Björk}, {and} \bibinfo{person}{Chris Lewis}.}
  \bibinfo{year}{2013}\natexlab{}.
\newblock \showarticletitle{Dark Patterns in the Design of Games}. In
  \bibinfo{booktitle}{\emph{Proceedings of the 8th International Conference on
  the Foundations of Digital Games (FDG 2013)}} (May 14-17).
  \bibinfo{publisher}{Society for the Advancement of the Science of Digital
  Games}, \bibinfo{address}{Chania, Crete, Greece}, \bibinfo{pages}{39--46}.
\newblock
\showISBNx{978-0-9913982-0-1}
\urldef\tempurl%
\url{http://www.fdg2013.org/program/papers.html}
\showURL{%
\tempurl}


\end{thebibliography}

\end{document}